\documentclass[11pt]{article}

\usepackage[margin=2.5cm]{geometry}
\usepackage[round]{natbib}
\usepackage{textcomp}

\usepackage[parfill]{parskip}

\usepackage[utf8]{inputenc}
\usepackage{hyperref}
\hypersetup{
     colorlinks   = true,
     citecolor    = blue,
     linkcolor 	= blue
}
\usepackage{amsmath,amsfonts}
\usepackage{amssymb}
\usepackage{amsthm}
\usepackage{graphicx}
\usepackage{caption}
\usepackage{sty/notation}
\usepackage{bm}
\usepackage{verbatim}
\usepackage{multirow}
\usepackage{hhline}
\usepackage[british]{babel}
\usepackage{pifont}
\newcommand{\cmark}{\ding{51}}%
\newcommand{\xmark}{\ding{55}}%
\usepackage{ctable}

\usepackage{array}
\newcolumntype{L}[1]{>{\raggedright\let\newline\\\arraybackslash\hspace{0pt}}m{#1}}

\usepackage{caption}
\usepackage{subcaption}

\newcommand{\postodds}[1]{\ensuremath{\text{\tt post-odds}(\; \H_1 \/\text{vs.}\/ \H_0 \; |\; #1)}}
\newcommand{\pjef}{\ensuremath{\mathbb P}_{\text{\sc j}}}
\renewcommand{\H}{\mathcal H}

\DeclareMathOperator*{\pr}{\mathbb P}

\DeclareBoldMathCommand{\A}{A}
\DeclareBoldMathCommand{\vmu}{\mu}
\DeclareBoldMathCommand{\zeros}{0}
\DeclareBoldMathCommand{\SSigma}{\Sigma}

\DeclareRobustCommand{\VANDER}[3]{#2}

\title{Why optional stopping can be a problem for Bayesians\footnote{We
    are grateful to Jeffrey Rouder, Eric-Jan Wagenmakers,
    Alexander Ly and the AE for providing important feedback on an earlier version of this
    article. Rouder and Wagenmakers both noticed significant
    unclarities and errors, which prompted us to do a substantial rewrite of
    the article.}  } \author{Rianne de Heide, Peter D. Gr{\"u}nwald}

\begin{document}
\maketitle

\begin{abstract}
  Recently, optional stopping has been a subject of debate in the
  Bayesian psychology community.  \citet{Rouder2014} argues that
  optional stopping is no problem for Bayesians, and even recommends
  the use of optional stopping in practice, as do
  \citet{Wagenmakers2012}. This article addresses the question whether
  optional stopping is problematic for Bayesian methods, and specifies
  under which circumstances and in which sense it is and is not. By
  slightly varying and extending Rouder's (\citeyear{Rouder2014})
  experiments, we illustrate that, as soon as the parameters of interest 
  are equipped with default or pragmatic priors --- which means, in
  most practical applications of Bayes factor hypothesis testing ---
  resilience to optional stopping can break down. We distinguish
  between three types of default priors, each having their own
  specific issues with optional stopping, ranging from no-problem-at-all (Type~0 priors) to quite severe (Type II priors).
\end{abstract}

\section{Introduction}
P-value based null-hypothesis significance testing (NHST) is widely
used in the life and behavioral sciences, even though the use of
$p$-values has been severely criticized for at least the
last 50 years.  During the last decade, within the field of
psychology, several authors have advocated the Bayes factor as the
most principled alternative to resolve the problems with
$p$-values. Subsequently, these authors have made an admirable effort
to provide practitioners with \emph{default Bayes factors} for common
hypothesis tests. Key references include, among many others, \citet{Rouder2009, Jamil2016, Rouder2012}.

We agree with the objections against the use of p-value based NHST and
the view that this paradigm is inappropriate (or at least far from
optimal) for scientific research, and we agree that the Bayes factor
has many advantages. However, as also noted by \cite{Gigerenzer2016}, it is
not the panacea for hypothesis testing that a lot of articles make it
appear. The Bayes factor has its limitations (cf.\ also \citet{tendeiro2019review}), and it seems that the
subtleties of when those limitations apply sometimes get lost in the
overwhelming effort to provide a solution to the pervasive problems of
p-values.

In this article we elucidate the intricacies of handling optional
stopping with Bayes factors, primarily in response to
\citet{Rouder2014}. \emph{Optional stopping} refers to `looking at the
results so far to decide whether or not to gather more data', and it
is a desirable property of a hypothesis test to be able to
\emph{handle optional stopping}. The key question is whether Bayes
factors can or cannot handle optional stopping. \citet{Yu2014},
\citet{Sanborn2014} and \citet{Rouder2014} tried to answer this
question from different perspectives and with different
interpretations of the notion of handling optional
  stopping. \citet{Rouder2014} illustrates, using computer
simulations, that optional stopping is not a problem for Bayesians,
also citing \citet{Lindley1957} and \citet{Edwards1963} who provide
mathematical results to a similar (but not exactly the same) effect. Rouder used the simulations to concretely illustrate more abstract mathematical theorems; these theorems are indeed formally proven by \citet{deng2016continuous} and, in a more general setting, by \citet{OptionalStoppingTechnical}. Other early work indicating
that optional stopping is not a problem for Bayesians includes
\citet{Savage1972} and \citet{Good1991}. We briefly return to all of these in Section~\ref{sec:other}. 

All this earlier work notwithstanding, we maintain that optional
stopping can be a problem for Bayesians --- at least for {\em
  pragmatic Bayesians\/} who are either  
willing to use so-called `default',
or `convenience' priors, or otherwise are willing to admit that their priors are imperfect and are willing to subject them to robustness analyses. In practice, nearly all   statisticians who use Bayesian methods are `pragmatic' in this sense.

\citet{Rouder2014} was written mainly in response to \citet{Yu2014},
and his main goal was to show that Bayesian procedures retain a clear
interpretation under optional stopping. He presents a criterion which,
if it holds for a given Bayesian method, indicates that, in some
specific sense, it performs as one would hope under optional
stopping. The main content of this article is to investigate this
criterion, which one may call {\em prior-based calibration}, for
common testing scenarios involving default priors. We shall encounter
two types of default priors, and we shall see that Rouder's
calibration criterion --- while indeed providing a clear {\em interpretation\/}
to Bayesian optional stopping whenever defined --- is in many cases either of limited {\em relevance\/}
(Type I priors) or {\em undefined\/} (Type II priors).

We consider a strengthening of Rouder's check which we call {\em
  strong calibration}, and which remains meaningful for all default
priors. Then, however, we shall see that strong calibration fails to
hold under optional stopping for all default priors except,
interestingly, for a special type of priors (which we call ``Type~0
priors'') on a special (but common) type of nuisance parameters. Since
these are rarely the only parameters incurring in one's models, one
has to conclude that optional stopping is usually a problem for
pragmatic Bayesians --- at least under Rouder's calibration criterion
of handling optional stopping. There exist (at least) two other
reasonable definitions of `handling optional stopping', 
which we provide in Section~\ref{sec:other}. There we also discuss
how, under these alternative definitions, Type I priors are sometimes
less problematic, but Type II priors still are.
As explained in the conclusion (Section~\ref{sec:discussion,
  conclusion}), the overall crux is that default and pragmatic priors
represent {\em tools\/} for inference just as much or even more than
{\em beliefs\/} about the world, and should thus be equipped with a
precise prescription as to what type of inferences they can and cannot
be used for. A first step towards implementing this radical idea is
given by one of us in the recent paper {\em Safe Probability\/}
\citep{Grunwald17}.

Readers who are familiar with Bayesian theory will not be too
surprised by our conclusions: It is well-known that what we call Type
II priors violate the {\em likelihood principle\/}
\citep{BergerWolpert1988} and/or lead to (mild) forms of {\em
  incoherence\/} \citep{Seidenfeld1979} and, because of the close
connection between these two concepts and optional stopping, it should
not be too surprising that issues 
arise. Yet it is still useful to show how these issues pan out in
simple computer simulations, especially given the apparently common
belief that optional stopping is {\em never\/} a problem for
Bayesians. The simulations will also serve to illustrate the
difference between the subjective, pragmatic and objective views of
Bayesian inference, a distinction which matters a lot and which, we
feel, has been underemphasized in the psychology literature --- our
simulations may in fact serve to help the reader decide what viewpoint
he or she likes best. 

In Section~\ref{sec:intro bayes} we explain important concepts of
Bayesianism and Bayes factors.  Section~\ref{sec:experiments}
explains Rouder's calibration criterion and repeats and extends Rouder's illustrative experiments, 
showing the sense in which optional stopping is indeed not a
problem for Bayesians. Section~\ref{sec:subj bayesian optional
  stopping} then contains additional simulations indicating the
problems with default priors as summarized above. In Section~\ref{sec:other} we discuss conceptualizations of `handling optional stopping' that are different from Rouder's; this includes an explication of the purely subjective Bayesian viewpoint as well as an explication of a 
frequentist treatment of handling
optional stopping, which only concerns sampling under the null
hypothesis. We illustrate that some (not all!) Bayes factor
methods can handle optional stopping in this frequentist
sense.  We conclude with a discussion of our findings in
Section~\ref{sec:discussion, conclusion}.
  
\section{Bayesian probability and Bayes factors}\label{sec:intro bayes}
Bayesianism is about a certain interpretation of the concept
\emph{probability}: as \emph{degrees of
  belief}. \citet{Wagenmakers2007} and \citet{Rouder2014} give an
intuitive explanation for the different views of frequentists and
Bayesians in statistics, on the basis of coin flips. The frequentists
interpret probability as a limiting frequency. Suppose we flip a coin
many times, if the probability of heads is $3/4$, we see a proportion
of $3/4$ of all those coin flips with heads up. Bayesians interpret
probability as a degree of belief. If an agent believes the
probability of heads is $3/4$, she believes that it will be $3$ times
more likely that the next coin flip will result in heads than tails;
we return to the operational meaning of such a `belief' in terms of
betting in Section~\ref{sec:other}.  

A Bayesian first expresses this belief as a probability function. In our coin flipping
example, it might be that the agent believes that it is more likely
that the coin is biased towards heads, which the probability function
thus reflects. We call this the \emph{prior distribution}, and we
denote\footnote{With some abuse of notation, we use $\pr$ both to
  denote a generic probability distribution (defined on sets), and to
  denote its associated probability mass function and a probability
  density function (defined on elements of sets); whenever in this
  article we write $\pr(z)$ where $z$ takes values in a real-valued
  scalar or vector space, this should be read as $f(z)$ where $f$ is
  the density of $\pr$.} it by $\pr(\theta)$, where $\theta$ is the
parameter (or several parameters) of the model. In our example,
$\theta$ expresses the bias of the coin, and is a real number between
$0$ and $1$. After the specification of the prior, we conduct the
experiment and obtain the data $D$ and the likelihood $\pr(D |
\theta)$. Now we can compute the \emph{posterior distribution}
$\pr(\theta | D)$ with the help of \emph{Bayes' theorem}:
\begin{align}\label{eq:Bayes theorem}
\pr(\theta | D) = \frac{\pr(D | \theta)\pr(\theta)}{\pr(D)}.
\end{align}
\citet{Rouder2014} and \citet{Wagenmakers2007} provide 
a clear explanation of Bayesian hypothesis testing with Bayes factors \citep{Jeffreys1961, Kass1995}, which we repeat here for completeness. Suppose we want to test a null hypothesis $\H_0$ against an alternative hypothesis $\H_1$. A hypothesis can consist of a single distribution, for example: `the coin is fair'. We call this a \emph{simple hypothesis}. A hypothesis can also consist of two or more, or even infinitely many hypotheses, which we call a \emph{composite hypothesis}. An example is: `the coin is biased towards heads', so the probability of heads can be any number between $0.5$ and $1$, and there are infinitely many of those numbers. Suppose again that we want to test $\H_0$ against $\H_1$. We start with the so called \emph{prior odds}: $\pr(\H_1) / \pr(\H_0)$, our belief before seeing the data. Let's say we believe that both hypotheses are equally probable, then our prior odds are $1$-to-$1$. Next we gather data $D$, and update our odds with the new knowledge, using Bayes' theorem (Eq.~\ref{eq:Bayes theorem}):
\begin{align}\label{eq: posterior odds, bayes rule}
\postodds{D} = \frac{\pr(\H_1 | D)}{\pr(\H_0 | D)} =  \frac{\pr(\H_1)}{\pr(\H_0)} \frac{\pr(D | \H_1)}{\pr(D | \H_0)}.
\end{align}
The left term is called \emph{posterior odds}, it is our updated
belief about which hypothesis is more likely. Right of the prior odds,
we see the \emph{Bayes factor}, the term that describes how the
beliefs (prior odds) are updated via the data. If we have no
preference for one hypothesis and set the prior odds to $1$-to-$1$, we
see that the posterior odds are just the Bayes factor. If we test a
composite $\H_0$ against a composite $\H_1$, the Bayes factor is a
ratio of two likelihoods in which we have two or more possible values
of our parameter $\theta$. Bayesian inference tells us how to
calculate $\pr(D \mid \H_j)$: we integrate out the parameter with help
of a prior distribution $\pr(\theta)$, and we write Eq.~\eqref{eq:
  posterior odds, bayes rule} as:
\begin{align}\label{eq:posterior odds, BF with integrals}
\postodds{D} = \frac{\pr(\H_1 | D)}{\pr(\H_0 | D)} =  \frac{\pr(\H_1)}{\pr(\H_0)} \frac{\int_{\theta_1} \pr(D | \theta_1) \pr (\theta_1) \dif \theta_1}{\int_{\theta_0} \pr(D | \theta_0) \pr (\theta_0) \dif \theta_0}
\end{align}
where $\theta_0$ denotes the parameter of the null hypothesis $\H_0$, and similarly, $\theta_1$ is the parameter of the alternative hypothesis $\H_1$. If we observe a Bayes factor of $10$, it means that the \emph{change} in odds from prior to posterior in favor of the alternative hypothesis $\H_1$ is a factor $10$. Intuitively, the Bayes factor provides a measure of whether the data have increased or decreased the odds on $\H_1$ relative to $\H_0$.

\section{Handling Optional stopping in the Calibration Sense}\label{sec:experiments}
Validity under optional stopping is a desirable property of hypothesis testing: we gather some data, look at the results, and decide whether we stop or gather some additional data. Informally we call `peeking at the results to decide whether to collect more data' \emph{optional stopping}, but if we want to make more precise what it means if we say that a test can handle optional stopping, it turns out that different approaches (frequentist, subjective Bayesian and objective Bayesian) lead to different interpretations or definitions. In this section we adopt the definition of handling optional stopping that was used by Rouder, and show, by repeating and extending Rouder's original simulation, that Bayesian methods do handle optional stopping in this sense. In the next section, we shall then see that for `default' and `pragmatic' priors used in practice, Rouder's original definition may not always be appropriate --- indicating there are problems with optional stopping after all.


\subsection{Example 0: Rouder's example}

We start by repeating Rouder's (\citeyear{Rouder2014}) second example,
so as to explain his ideas and re-state his results. Suppose a
researcher wants to test the null hypothesis $\H_0$ that the mean of a
normal distribution is equal to $0$, against the alternative
hypothesis $\H_1$ that the mean is not $0$: we are really testing whether $\mu=0$ or not. In Bayesian statistics,
the  composite alternative $\H_1: \mu \neq 0$ is incomplete without specifying a prior
on $\mu$; like in Rouder's example, we take the prior on the mean
to be a standard normal, which is a fairly standard (though by no means the only common) choice \citep{Berger85,BernardoS94}. This expresses a belief that small effect
sizes are possible (though the prior probability of the mean being {\em
  exactly\/} 0 is $0$), while a mean as large as $1.0$ is neither
typical nor exceedingly rare. We take the variance to be $1$, such
that the mean  equals the effect size. We set our prior
odds to $1$-to-$1$: This expresses a priori indifference between the
hypotheses, or a belief that both hypotheses are really equally
probable.
To give a first example, suppose we observe 
$n=10$ observations
Now we can observe the
data and update our prior beliefs. We calculate the posterior odds, in our case equal to the Bayes factor, via
Eq.~\eqref{eq: posterior odds, bayes rule} for data $D = (x_1, \ldots, x_n)$:
\begin{align} \label{eq:posterior odds Rouders experiment}
  \postodds{x_1,\ldots,x_n} &= \frac{1}{1} \cdot \frac{\exp\left\lbrace
      \frac{n^2 \overline{x}^2}{2(n+1)} \right\rbrace}{\sqrt{n+1}}
\end{align}
\noindent where $n$ is the sample size ($10$ in our case), and $\overline{x}$ is the sample mean.
Suppose we observe posterior odds of $3.5$-to-$1$ in favor of the null. 
\paragraph{Calibration, Mathematically}
As Rouder writes: `If a replicate experiment yielded a posterior odds of $3.5$-to-$1$ in favor of the null, then we expect that the null was $3.5$ times as probable as the alternative to have produced the data.' In mathematical language, this can be expressed as
\begin{equation}\label{eq:calibrate}
\postodds{\text{``$\postodds{x_1,\ldots, x_n} = a$''}} \  =\  a,
\end{equation}
for the specific case $n=10$ and $a=1/3.5$; of course we would expect
this to hold for general $n$ and $a$.  The quotation marks indicate
that we condition on an event, i.e.\ a set of different data realizations; in our case this is the set  of all data $x_1, \ldots, x_n$ for which the posterior odds are $a$.  We say that
(\ref{eq:calibrate}) expresses {\em calibration of the posterior
  odds}. To explain further, we draw the analogy to weather
forecasting: consider a weather forecaster who, on each day, announces
the probability that it will rain the next day at a certain
location. It is standard terminology to call such a weather forecaster
{\em calibrated\/} if, on average on those days for which he predicts
`probability of rain is $30\%$', it rains about $30\%$ of the time, on
those days for which he predicts $40\%$, it rains $40\%$ of the time,
and so on. Thus, although his predictions presumably depend on a lot
of data such as temperature, air pressure at various locations etc.,
given {\em only\/} the fact that this data was such that he predicts $a$, the
actual probability is $a$. Similarly, given only the fact the
posterior odds based on the full data are $a$ (but not given the full
data itself), the posterior odds should still be $a$ (readers who find
(\ref{eq:calibrate}) hard to interpret are urged to study the
simulations below).

Indeed, it turns out that  (\ref{eq:calibrate}) is the case. This can be shown either as a mathematical theorem, or, as Rouder does, by computer simulation.  At this point, the result is merely a sanity check, telling us that Bayesian updating is not crazy, and is not really surprising.
Now, instead of a fixed $n$, let us consider optional stopping: we
keep adding data points until the the posterior odds are at least $10$-to-$1$
for either hypothesis, unless a maximum of $25$ data points was reached. Let $\tau$ be the sample size (which is now data-dependent) at which we stop; note that $\tau \leq 25$. Remarkably, it turns out that we still have
\begin{equation}\label{eq:calibrateb}
\postodds{\text{``$\postodds{x_1,\ldots, x_{\tau}} = a$''}} \  =\  a,
\end{equation}
for this (and in fact any other data-dependent) stopping time $\tau$. In words, {\em the posterior odds remain calibrated under optional stopping}. Again, this can be shown formally, as a mathematical theorem (we do so in \cite{OptionalStoppingTechnical}; see also \cite{deng2016continuous}). 

\paragraph{Calibration, Proof by Simulation}
Following \citet{Yu2014} and \citet{Sanborn2014}, Rouder uses
computer simulations, rather than mathematical derivation, to
elucidate the properties of analytic methods. In Rouder's words `this
choice is wise for a readership of experimental
psychologists. Simulation results have a tangible, experimental feel;
moreover, if something is true mathematically, we should be able to
see it in simulation as well'.  Rouder illustrates both
(\ref{eq:calibrate}) and (\ref{eq:calibrateb}) by a simulation which
we now describe.

\begin{figure}
    \centering
    \begin{subfigure}[b]{0.5\textwidth}
        \includegraphics[width=\textwidth]{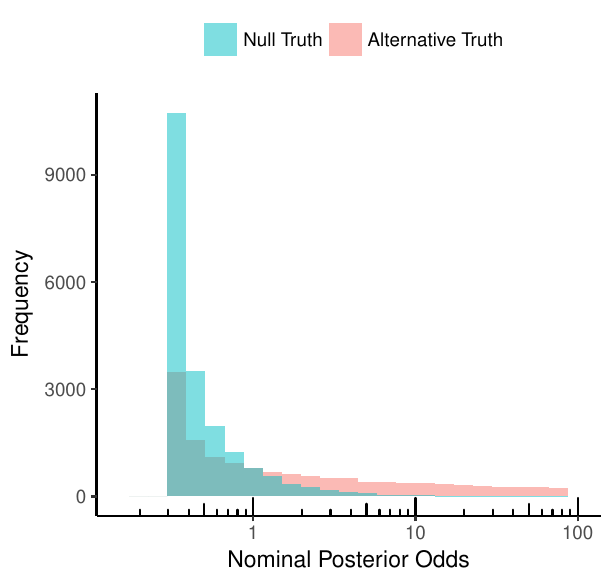}
        \caption{}
        \label{fig:Rouder a}
    \end{subfigure}
    \hfill
    \begin{subfigure}[b]{0.4\textwidth}
        \includegraphics[width=\textwidth]{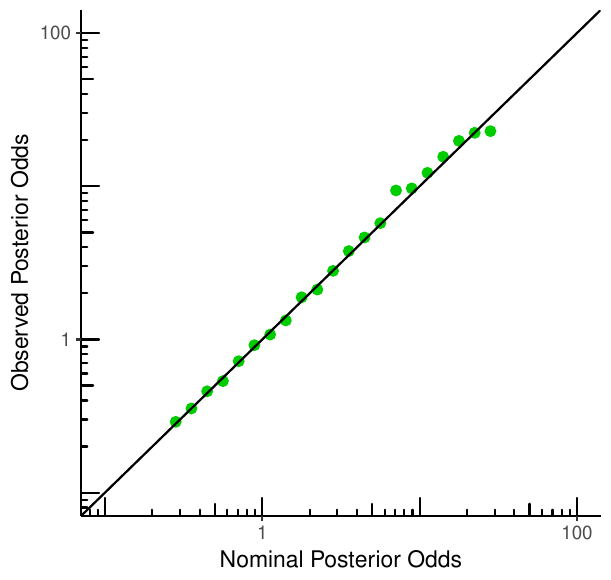}
        \caption{}
        \label{fig:Rouder b}
    \end{subfigure}
\caption{The interpretation of the posterior odds in Rouder's experiment, from $20,000$ replicate experiments. (a) The empirical sampling distribution of the posterior odds as a histogram under $\H_0$ and $\H_1$. (b) Calibration plot: the observed posterior odds as a function of the nominal posterior odds. }\label{fig:plaatjes Rouder}
\end{figure}

Again we draw data from the null hypothesis: say $n=10$
observations from a normal distribution with mean $0$ and variance
$1$.  But now we repeat this procedure $20,000$ times, and we see the
distribution of the posterior odds plotted as the blue histogram on
the log scale in Figure~\ref{fig:Rouder a}. We also sample data from
the alternative distribution $\H_1$: first we sample a mean from a
standard normal distribution (readers that consider this `sampling from the prior' to be strange are urged to read on), and then we sample $10$ observations
from a normal distribution with this just obtained mean, and variance
$1$. Next, we calculate the posterior odds from
Eq.~\eqref{eq:posterior odds Rouders experiment}. Again, we perform
$20,000$ replicate experiments of $10$ data points each, and we obtain
the pink histogram in Figure~\ref{fig:Rouder a}. We see that for the
null hypothesis, most samples favor the null (the values of the Bayes
factor are smaller than $1$), for the alternative hypothesis we see
that the bins for higher values of the posterior odds are higher.

In terms of this simulation, Rouder's claim that,
`If a replicate experiment yielded a posterior odds of $3.5$-to-$1$ in favor of the null, then we expect that the null was $3.5$ times as probable as the alternative to have produced the data',
as formalized by (\ref{eq:calibrate}),  now says the following: if we look at a specific bin of the histogram, say at $3.5$, i.e.\ the number of all the replicate experiments that yielded approximately a posterior odds of $3.5$, then the bin from $\H_1$ should be about $3.5$ times as high as the bin from $\H_0$. Rouder calls the ratio of the two histograms the \emph{observed posterior odds}: the ratio of the binned posterior odds counts we observe from the simulation experiments we did. What we expect the ratio to be for a certain value of the posterior odds, is what he calls the \emph{nominal posterior odds}. We can plot the observed posterior odds as a function of the nominal posterior odds, and we see the result in Figure~\ref{fig:Rouder b}. The observed values agree closely with the nominal values: all points lie within simulation error on the identity line, which can be considered as a `proof of (\ref{eq:calibrate}) by simulation'.

\citet{Rouder2014} repeats this experiment under optional stopping: he
ran a simulation experiment with exactly the same setup, except that
in each of the $40,000$ simulations, sampling occurred until the
posterior odds were at least $10$-to-$1$ for either hypothesis, unless
a maximum of $25$ observations was reached.  This yielded a figure
indistinguishable from Figure~\ref{fig:Rouder b}, from which Rouder
concluded that `the interpretation of the posterior odds holds with
optional stopping'; in our language, {\em the posterior odds remain
  calibrated under optional stopping} --- it is a proof, by simulation, that (\ref{eq:calibrateb}) holds. From this and similar
experiments, Rouder concluded that Bayes factors still have a clear
interpretation under optional stopping (we agree with this for what we
call below Type 0 and I priors, not Type II), leading to the
claim/title `optional stopping is no problem for Bayesians' (for which
we only agree for Type 0 and purely subjective priors).

\paragraph{Is sampling from the prior meaningful?} When presenting
Rouder's simulations to other researchers, a common concern is: `how
can sampling a parameter from the prior in ${\cal H}_1$ be meaningful?
In any real-life experiment, there is just one, fixed population
value, i.e.\ one fixed value of the parameter that governs the data.'
This is indeed true, and not in contradiction with Bayesian ideas:
Bayesian statisticians put a distribution on parameters in
${\cal H}_1$ that expresses their uncertainty about the parameter,
and that should not be interpreted as something that is `sampled'
from. Nevertheless, Bayesian posterior odds calculations are done by
calculating weighted averages via integrals, and the results are {\em
  mathematically equivalent\/} to what one gets if, as above, one
samples a parameter from the prior, and the data from the parameter,
and then takes averages over many repetitions. We (and Rouder) really
want to establish (\ref{eq:calibrate}) and (\ref{eq:calibrateb})
(which can be interpreted without resorting to sampling a parameter
from a prior), and we note that it is equivalent to the curve in
Figure~\ref{fig:Rouder b} coinciding with the diagonal.

Some readers of an earlier draft of this paper concluded that, given
its equivalence to an experiment involving sampling from the prior,
which feels meaningless to them, (\ref{eq:calibrateb}) is itself
invariably meaningless. Instead, they claim, because in real-life the parameter often has one specific fixed value, one should look at what
happens under sampling under fixed parameter values. Below we
shall see that if we look at such {\em strong calibration}, we
sometimes (Example 1) still get calibration, but usually (Example 2)
we do not; so such readers will likely agree with our conclusion that
`optional stopping can be a problem for Bayesians', even though they
would disagree with us on some details, because we do think that
(\ref{eq:calibrateb}) can be a meaningful statement for some, but not
all priors. To us, the importance of the simulations is simply to
verify (\ref{eq:calibrateb}) and, later on (Example 2), to show that
(\ref{eq:calibratec}), the stronger analogue of (\ref{eq:calibrateb})
that we would like to hold for default priors, does not always hold.

\subsection{Example 1: Rouder's example with a nuisance parameter}\label{sec:Rouders example with a nuisance parameter}
We now adjust Rouder's example to a case where we still want to test
whether $\mu=0$, but the variance $\sigma^2$ is unknown. Posterior
calibration will still be obtained under optional stopping; the
example mainly serves to gently introduce the notions of {\em improper
  prior\/} and {\em strong vs.\ prior calibration}, that will play a
central role later on. So, ${\cal H}_0$ now expresses that the data
are independently normally distributed with mean $0$ and some unknown
variance $\sigma^2$, and ${\cal H}_1$ expresses that the data are
normal with variance $\sigma^2$, and some mean $\mu$, where the
uncertainty about $\mu$ is once again captured by a normal prior: the
mean is distributed according to a normal  with mean zero
and variance (again) $\sigma^2$ (this corresponds to a standard normal
distribution on the effect size). If $\sigma^2 = 1$, this reduces to
Rouder's example; but we now allow for arbitrary $\sigma^2$.  We call
$\sigma^2$ a {\em nuisance parameter\/}: a parameter that occurs in
both models, is not directly of interest, but that needs to be
accounted for in the analysis. The setup is analogous to the standard 1-sample frequentist $t$-test, where we also want to test whether a mean is $0$ or not, without knowing the variance; in the Bayesian approach, such a test only becomes defined once we have a prior for the parameters. For $\mu$  we choose a normal,\footnote{The advantage of a normal is that it makes calculations relatively easy. A more common and perhaps more defensible choice is a Cauchy distribution, used in the `default Bayesian $t$-test', which we consider further below.} 
for the nuisance parameter $\sigma$ we will make the standard choice of  Jeffreys' prior for the variance:
$\pjef(\sigma) := 1/\sigma$ \citep{Rouder2009}. To obtain the Bayes
factor for this problem, we integrate out the parameter $\sigma$ cf.\
Eq.~(\ref{eq:posterior odds, BF with integrals}). Again, we assign
prior odds of $1$-to-$1$, and obtain the posterior odds:
\begin{align}\nonumber
\postodds{D} &= \frac{1}{1} \frac{ \int_0^\infty \frac{1}{\sigma} \prod_{i=1}^n \frac{1}{\sqrt{2\pi\sigma^2}} \exp\left(-\frac{x_i^2}{2\sigma^2} \right) \dif\sigma }{\int_0^\infty \frac{1}{\sigma} \int_{-\infty}^{\infty} \frac{1}{\sqrt{2\pi\sigma^2}} \exp\left( -\frac{\mu}{2\sigma^2} \right) \prod_{i=1}^n \frac{1}{\sqrt{2\pi\sigma^2}} \exp\left(-\frac{(x_i-\mu)^2}{2\sigma^2} \right) \dif\mu \dif\sigma} \\ \nonumber
&= \frac{1}{\sqrt{n+1}}\left(1 - \frac{\left( \frac{1}{n+1}\sum_{i=1}^n x_i \right)^2}{\frac{1}{n+1}\sum_{i=1}^n x_i^2} \right)^{-\frac{n}{2}}
\end{align}
Formally, Jeffreys' prior on $\sigma$ is a
`measure' rather than a distribution, since it does not integrate to $1$: clearly 
\begin{align}\label{eq:Jeffreys prior sigma normaal}
\int_{0}^{\infty} \pjef(\sigma) \dif \sigma =  \int_0^{\infty} \frac{1}{\sigma} \dif \sigma  = \infty,
\end{align}
Priors that integrate to
infinity are often called \emph{improper}. Use of such priors for nuisance parameters is not
really a problem for Bayesian inference, since one can typically plug such
priors into Bayes' theorem anyway, and this leads to proper
posteriors, i.e.\ posteriors that do integrate to one, and then the
Bayesian machinery can go ahead. Since  Jeffreys' prior is meant to express that we have no clear prior knowledge about the variance, we would hope that Bayes would remain interpretable under optional stopping, no matter what the (unobservable) variance in our sampling distribution actually is. Remarkably, this is indeed the case: for all $\sigma_0^2 > 0$, we have the following analogue of (\ref{eq:calibrateb}): 
\begin{equation}\label{eq:calibratec}
\postodds{\sigma^2 = \sigma^2_0, \text{``$\postodds{x_1,\ldots, x_{\tau}} = a$''}}  \  =\  a,
\end{equation}
In words, this means that, given that the posterior odds (calculated based on Jeffreys' prior, i.e.\ without knowing the variance) are equal to $a$ {\em and\/} that the actual variance is $\sigma^2_0$, the posterior odds are still $a$, irrespective of what $\sigma^2_0$ actually is. This statement may be quite hard to interpret, so we proceed to illustrate it by simulation again. 

\begin{figure}
    \centering
    \begin{subfigure}[b]{0.4\textwidth}
        \includegraphics[width=\textwidth]{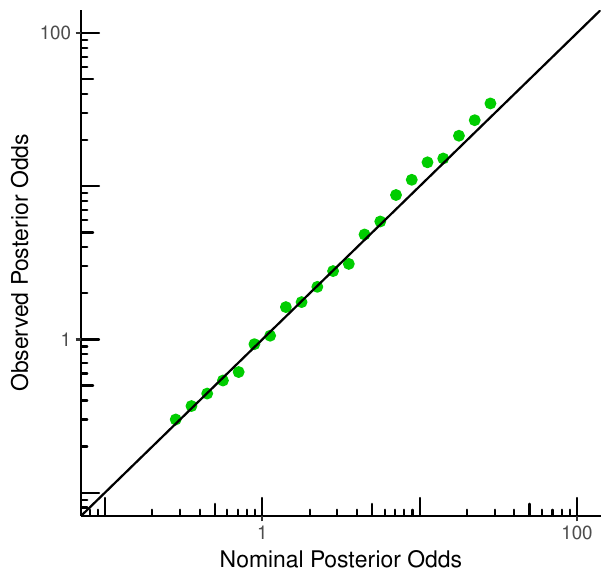}
        \caption{}
        \label{fig:RouderExtra a}
    \end{subfigure}
    \hfill
    \begin{subfigure}[b]{0.4\textwidth}
        \includegraphics[width=\textwidth]{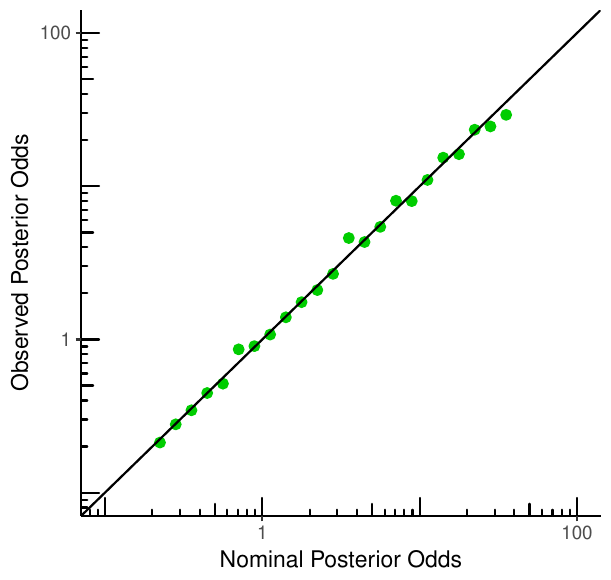}
        \caption{}
        \label{fig:RouderExtra b}
    \end{subfigure}
    \caption{Calibration of the experiment of Section~\ref{sec:Rouders example with a nuisance parameter}, from $20,000$ replicate experiments. (a) The observed posterior odds as a function of the nominal posterior odds. (b) The observed posterior odds as a function of the nominal posterior odds with optional stopping.}\label{fig:plaatjes RouderExtra}
\end{figure}

To repeat Rouder's experiment, we have to simulate data under both
${\cal H}_0$ and ${\cal H}_1$. To do this we need to specify the
variance $\sigma^2$ of the normal distribution(s) from which we
sample. Whereas, as in the previous experiment, we can sample the mean in ${\cal H}_1$ from the prior, for the variance we seem to run into a problem: it is not clear how one should sample from an improper prior. 
$\theta$. But we cannot directly sample $\sigma$ from an improper prior.
As an alternative,
we can pick any particular fixed $\sigma^2$ to sample from, as we now illustrate. Let us first try $\sigma^2 = 1$. 
Like Rouder's example, we sample the mean of the alternative
hypothesis $\H_1$ from the aforementioned normal distribution. Then,
we sample $10$ data points from a normal distribution with the just
sampled mean and the variance that we picked. For the null hypothesis
$\H_0$ we sample the data from a normal distribution with mean zero
and the same variance. We continue the experiment just as Rouder did:
we calculate the posterior odds from $20,000$ replicate experiments of
$10$ generated observations for each hypothesis, and construct the histograms and the
plot of the ratio of the counts to see if calibration is violated. In
Figure \ref{fig:RouderExtra a} we see the calibration plot for the
experiment described above. In Figure \ref{fig:RouderExtra b} we see
the results for the same experiment, except that we performed optional
stopping: we sampled until the posterior odds were at least $10$-to-$1$
for $\H_1$, or the maximum of $25$ observations was reached. We see that
the posterior odds in this experiment with optional stopping are
calibrated as well. 

\paragraph{Prior Calibration vs.\ Strong Calibration}
Importantly, the same conclusion remains valid whether we sample data
using $\sigma^2 =1$, or $\sigma^2 = 2$, or any other value --- in
simulation terms (\ref{eq:calibratec}) simply expresses that we get
calibration (i.e.\ all points on the diagonal) no matter what
$\sigma^2$ we actually sample from: even though calculation of the
posterior odds given a sample makes use of the prior
$\pjef(\sigma) = 1/\sigma$ and does not know the `true' $\sigma$,
calibration is retained under sampling under arbitrary `true'
$\sigma$. We say that the posterior odds are {\em prior-calibrated\/}
for parameter $\mu$ and {\em strongly calibrated\/} for
$\sigma^2$. More generally and formally, consider general hypotheses
$\H_0$ and $\H_1$ (not necessarily expressing that data are normal)
that share parameters $\gamma_0, \gamma_1$ and suppose that
(\ref{eq:calibratec}) holds with $\gamma_1$ in the role of
$\sigma^2$. Then we say that $\gamma_0$ is prior-calibrated (to get
calibration in simulations we need to draw it from the prior) and
$\gamma_1$ is strongly calibrated (calibration is obtained when
drawing data under all possible $\gamma_1$).

Notably, strong calibration is a special property of the chosen prior.
if we had chosen another proper or improper prior to calculate the
posterior odds 
(for example, the improper prior ${\mathbb
  P}'(\sigma) \propto \sigma^{-2}$ has sometimes been used in this
context) then the property that calibration under optional stopping is
retained under any choice of $\sigma^2$ will cease to hold; we will
see examples below. The reason that $\pjef(\sigma) \propto 1/\sigma$
has this nice property is that $\sigma$ is a special type of nuisance
parameter for which there exists a suitable group structure, relative
to which both models are invariant
\citep{Eaton1989,Berger1998,Berger2003}. This sounds more complicated
than it is --- in our example, the invariance is scale invariance: if
we divide all outcomes by any fixed $\sigma$ (multiply by $1/\sigma$),
then the Bayes factor remains unchanged; similarly, one may have for
example location invariances.

If such group structure parameters are equipped with a special
prior (which, for reasons to become clear, we shall term {\em Type~0
  prior\/}), then we obtain strong calibration, both for fixed sample
sizes and under optional stopping, relative to these
parameters.\footnote{Technically, the Type~0 prior for a given group
  structure is defined as the right-Haar prior for the group \citep{Berger1998}: a unique
  (up to a constant) probability measure induced on the parameter
  space by the right Haar measure on the related group. Strong
  calibration is proven in general by \cite{OptionalStoppingTechnical}, and \cite{Hendriksen2017} for the special case of the $1$-sample $t$-test.} Jeffreys' prior for the variance
$\pjef(\sigma)$ is the Type~0 prior for the variance nuisance
parameter. \citet{Berger2003} show that such priors can be defined for
a large class of nuisance parameters --- we will see the example of a
prior on a common mean rather than a variance in Example 3 below; but
there also exist cases with parameters that (at least intuitively) are
nuisance parameters, for which Type~0 priors do not exist; we give an
example in the Appendix. For parameters
of interest, including e.g.\ any parameter that does not occur in both
models, Type~0 priors never exist. Thus, strong calibration cannot be obtained for those parameters.

\section{When Problems arise: Subjective vs.\ Pragmatic and Default Priors}
\label{sec:subj bayesian optional stopping}
%
Bayesians view probabilities as degree of belief. The degree of belief
an agent has before conducting the experiment, is expressed as a
probability function. This \emph{prior} is then updated with data
from experiments, and the resulting \emph{posterior} can be used
to base decisions on. 

For one pole of the spectrum of Bayesians, the
pure \emph{subjectivists}, this is the full story
\citep{DeFinetti1937, Savage1972}: any prior capturing the belief of
the agent is allowed, but it should always be interpreted as the
agent's personal degree of belief; in Section~\ref{sec:other} we explain what such a `belief' really means. 

On the other end of the spectrum, the \emph{objective
  Bayesians} \citep{Jeffreys1961,Berger2006} argue that {degrees
  of belief} should be restricted, ideally in such a way that they
do not depend on the agent, and in the extreme case boil down to a
single, rational, probability function, where a priori distributions
represent indifference rather than subjective belief and a posteriori
distributions represent `rational degrees of confirmation' rather than
subjective belief. Ideally, in any given situation there should then
just be a single appropriate prior. Most objective Bayesians do not
take such an extreme stance, recommending instead {\em default\/}
priors to be used whenever only very little a priori knowledge is available.  These make a \emph{default} choice for the functional form of a distribution (e.g.\ Cauchy) but often have one or two parameters that can be specified in a subjective way. 
These may then be replaced by more informative priors when more
knowledge becomes available after all. We will see several examples of such
default priors below.

So what category of priors is used in practice?  Recent papers that
advocate the use of Bayesian methods within psychology such as
\citet{Rouder2009,Rouder2012,Jamil2016} are mostly based on default
priors. Within the statistics community, nowadays a pragmatic stance
is by far the most common, in which priors are used that mix `default'
and `subjective' aspects \citep{Gelman17} and that are also chosen to
allow for computationally feasible inference. Very broadly speaking,
we may say that there is a scale ranging from completely `objective'
(and hardly used) via `default' (with a few, say 1 or 2 parameters to
be filled in subjectively, i.e.\ based on prior knowledge) and
`pragmatic' (with functional forms of the prior based partly on prior
knowledge, partly by defaults, and partly by convenience) to the fully
subjective. Within the pragmatic stance, one explicitly acknowledges
that one's prior distribution may have some arbitrary aspects to it
(e.g.\ chosen to make computations easier rather than reflecting true
prior knowledge). It then becomes important to do sensitivity
analyses: studying what happens if a modified prior is used or if data
are sampled not by first sampling parameters $\theta$ from the prior
and then data from ${\mathbb P}(\cdot \mid\theta)$ but rather directly
from a fixed $\theta$ within a region that does not have overly small
prior probability.\footnote{To witness, one of us recently spoke at
  the bi-annual OBAYES (Objective Bayes) conference, and noticed that
  a substantial fraction of the talks featured such fixed
  $\theta$-analyses and/or used priors of Type II below.}

The point of this article is that Rouder's view on what constitutes `handling optional stopping' is tailored
towards a fully subjective interpretation of Bayes; as soon as one allows default and
pragmatic priors, problems with optional stopping do occur (except for what we call Type~0 priors). We can
distinguish between three types of problems, depending on the type of
prior that is used. We now give an overview of type of prior and
problem, giving concrete examples later.
\begin{enumerate}
\item{\em Type 0 Priors:\/} these are priors on parameters freely
  occurring in both hypotheses for which strong calibration (as with
  $\sigma^2$ in (\ref{eq:calibratec})) holds under optional
  stopping. This includes all right Haar priors on parameters that satisfy a group structure; \citet{OptionalStoppingTechnical} give a formal definition; \citet{Berger2003,Berger1998} give an overview of such priors.
We conjecture, but have no proof, that such right Haar priors on group structure parameters are the {\em only\/} priors allowing for strong calibration under optional stopping, i.e.\ the only Type 0 Priors. Some, but not all so-called `nuisance parameters' admit group structure/right Haar priors. For example, the variance in the $t$-test setting does, but the mean in $2\times2$ contingency tables \citep{OptionalStopping} does not. 
\item {\em Type I Priors:\/} these are default or pragmatic priors
  that do {\em not\/} depend on any aspects of the experimental setup
  (such as the sample size) or the data (such as the values of
  covariates) and are not of Type 0 above. Thus, strong calibration
  under optional stopping is violated with such priors --- an example
  is the Cauchy prior in Example 2 of Section~\ref{sec:Bayesian t test
    with default priors} below.

\item {\em Type II Priors:\/} these are default and pragmatic  priors that are not of Type 0 or I: the priors may themselves depend on the experimental setup, such as the sample size, the covariates (design), or the stopping time itself, or other aspects of the data. Such priors are quite common in the Bayesian literature. Here the problem is more serious: as we shall see, prior calibration is ill-defined, and correspondingly  Rouder's experiments cannot be performed for such priors, and `handling optional stopping' is in a sense impossible in principle. An example is the $g$-prior for regression as in Example 3 below or Jeffreys' prior for the Bernoulli model as in Section~\ref{sec:bernoulli} below. 
\end{enumerate}
We illustrate the problems with Type I and Type II priors by further
extending Rouder's experiment to two extensions of our earlier
setting, namely the Bayesian $t$-test, going back to \citet{Jeffreys1961}
and advocated by \citet{Rouder2009}, and objective Bayesian linear
regression, following \citet{Liang2008}. Both methods
are quite popular and use default Bayes factors based on default
priors, to be used when no clear or very little prior knowledge is readily available.

\subsection{Example 2: Bayesian $t$-test --- The Problem with Type I Priors}\label{sec:Bayesian t test with default priors}
Suppose a researcher wants to test the effect of a new fertilizer on
the growth of some wheat variety. The null hypothesis $\H_0$ states
that there is no difference between the old and the new
fertilizer, and the alternative hypothesis $\H_1$ states that the
fertilizers have a different effect on the growth of the wheat. We assume that
the length of the wheat is normally
distributed with the same (unknown) variance under both fertilizers,
and that with the old fertilizer, the mean is known to be $\mu_0=1$ meter.
We now take a number of seeds and apply the new fertilizer to each
of them. We
let the wheat grow for a couple of weeks, and  we measure the
lengths. The null
hypothesis $\H_0$ is thus: $\mu  = \mu_0 = 1$, and the alternative hypothesis
$\H_1$ is that the mean of the group with the new fertilizer is different
from $1$ meter: $\mu \neq 1$.

Again we follow Rouder's calibration check; again, the end goal is to illustrate a mathematical result, (\ref{eq:calibrated}) below, which will be contrasted with  (\ref{eq:calibrateb}). And again, to make the result concrete, we will first perform a simulation, generating data from both models
and updating our prior beliefs from this data as before. We do this using the
\emph{Bayesian $t$-test}, where Jeffreys' prior $\pjef(\sigma)=
1/\sigma$ is placed on the standard deviation $\sigma$ within both
hypotheses $\H_0$ and $\H_1$. Within $\H_0$ we set the mean to $\mu_0=1$ and within $\H_1$, a standard Cauchy
prior is placed on the effect size
$(\mu - \mu_0) / \sigma$; details are provided by \citet{Rouder2009}. Once
again, the nuisance parameter $\sigma$ is equipped with an improper
Jeffreys' prior, so, like in Experiment 1 above and for the reasons
detailed there, for simulating our data, we will choose a fixed value
for $\sigma$; the experiments will give the same result regardless
of the value we choose.

\begin{figure}
    \centering
    \begin{subfigure}[b]{0.5\textwidth}
        \includegraphics[width=\textwidth]{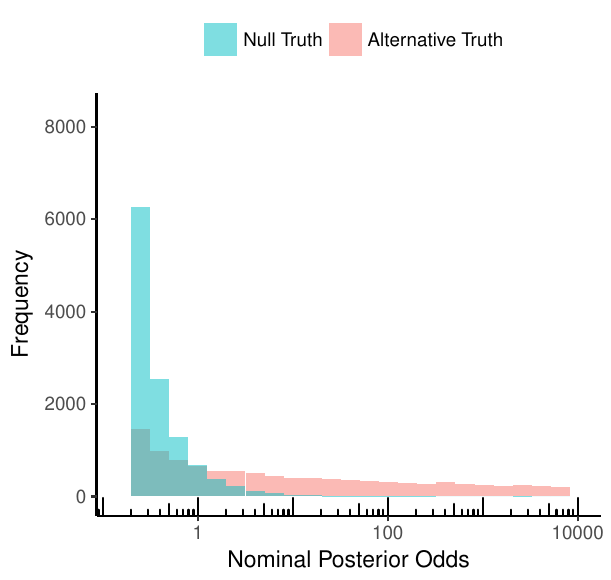}
        \caption{}
        \label{fig:ttest a}
    \end{subfigure}
    \hfill
    ~ 
    \begin{subfigure}[b]{0.4\textwidth}
        \includegraphics[width=\textwidth]{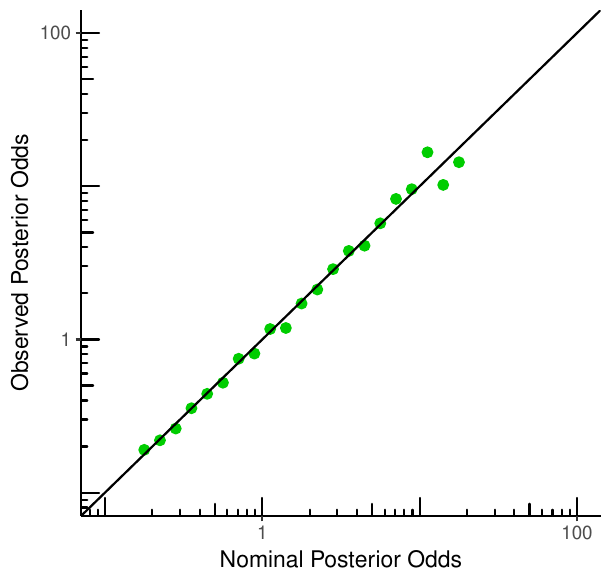}
        \caption{}
        \label{fig:ttest b}
    \end{subfigure}
    \vskip\baselineskip
        \begin{subfigure}[b]{0.5\textwidth}
        \includegraphics[width=\textwidth]{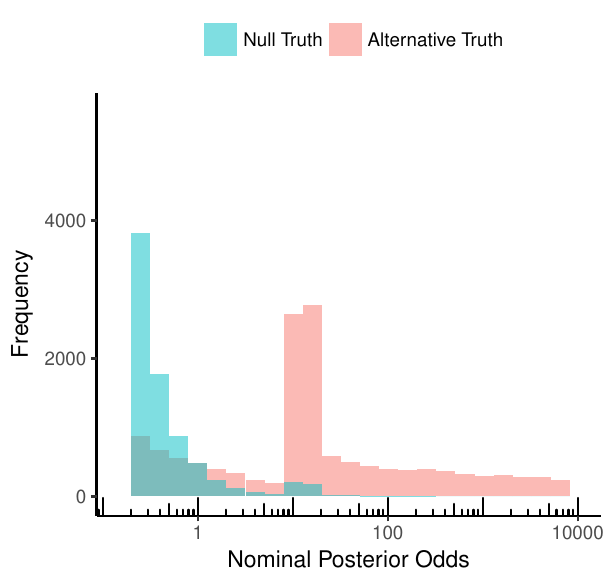}
        \caption{}
        \label{fig:ttest c}
    \end{subfigure}
    \hfill
    \begin{subfigure}[b]{0.4\textwidth}
        \includegraphics[width=\textwidth]{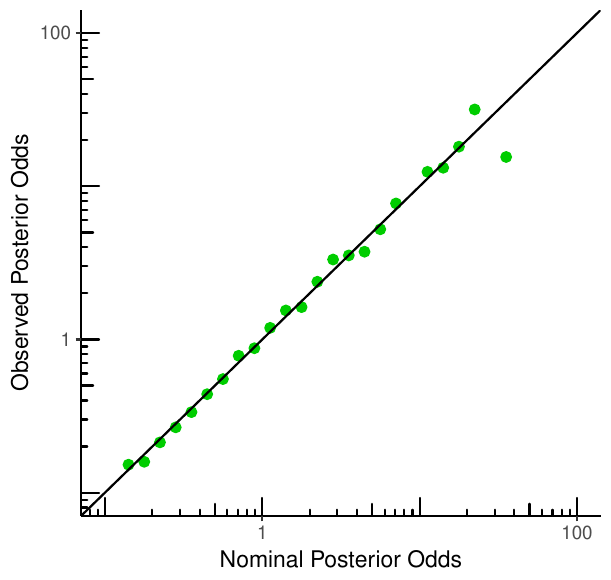}
        \caption{}
        \label{fig:ttest d}
    \end{subfigure}
    \caption{Calibration in the $t$-test experiment, Section~\ref{sec:Bayesian t test with default priors}, from $20,000$ replicate experiments. (a) The distribution of posterior odds as a histogram under $\H_0$ and $\H_1$ in one figure. (b) The observed posterior odds as a function of the nominal posterior odds. (c) Distribution of the posterior odds with optional stopping. (d) The observed posterior odds as a function of the nominal posterior odds with optional stopping.}\label{fig:plaatjes ttest}
\end{figure}

We generate $10$ observations for each fertilizer under both models:
for $\H_0$ we sample data from a normal distribution with mean
$\mu_0=1$ meter and we pick the variance $\sigma^2 = 1$. For $\H_1$ we
sample data from a normal distribution where the variance is $1$ as
well, and the mean is determined by the effect size above. We adopt a
Cauchy prior to express our beliefs about what values of the effect
size are likely, which is mathematically equivalent to the effect size
being sampled from a standard Cauchy distribution.  We follow Rouder's
experiment further, and set our prior odds on $\H_0$ and $\H_1$,
before observing the data, to $1$-to-$1$. 
We sample $10$ data points from each of the hypotheses, and
we calculate the Bayes factors. We repeat this procedure $20,000$ times. Then, we bin the $20,000$ resulting Bayes factors  and
construct a histogram. In Figure~\ref{fig:ttest a} we see the
distribution of the posterior odds when either the null or the
alternative are true in one figure. In Figure~\ref{fig:ttest b} we see
the calibration plot for this data from which Rouder checks the
interpretation of the posterior odds: the observed posterior odds is
the ratio of the two histograms, where the width of the bins is $0.1$
on the log scale. The posterior odds are calibrated, in accordance
with Rouder's experiments. We repeated the experiment with the
difference that in each of the $40,000$ experiments we sampled more
data points until the posterior odds were at least $10$-to-$1$, or the
maximum number of $25$ data points was reached. The histograms for this
experiment are in Figure~\ref{fig:ttest c}. In Figure~\ref{fig:ttest
  d} we can see that, as expected, the posterior odds are calibrated
under optional stopping
as well.

Since $\sigma^2$ is a nuisance parameter equipped with its Type~0 prior, it
does not matter what value we take when sampling data. We may ask
ourselves what happens if, similarly, we fix particular values of the
mean  and sample from them, rather than from the
prior; for sampling from $\H_0$, this does not change anything since the prior is concentrated on the single point $\mu_0 = 1$;  in $\H_1$, this means we can basically pick any $\mu$ and sample from it. In other words, we will check whether we have strong
calibration rather than prior-calibration not just for
$\sigma^2$, but also for the mean $\mu$. We now first describe such an
experiment, and will explain its importance further below. 

\begin{figure}
    \centering
    \begin{subfigure}[b]{0.4\textwidth}
        \includegraphics[width=\textwidth]{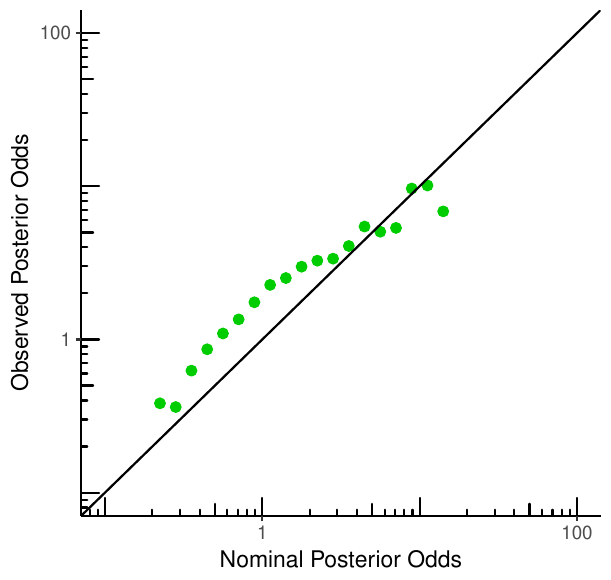}
        \caption{}
        \label{fig:TTestNonSubj a}
    \end{subfigure}
    \hfill
    \begin{subfigure}[b]{0.4\textwidth}
        \includegraphics[width=\textwidth]{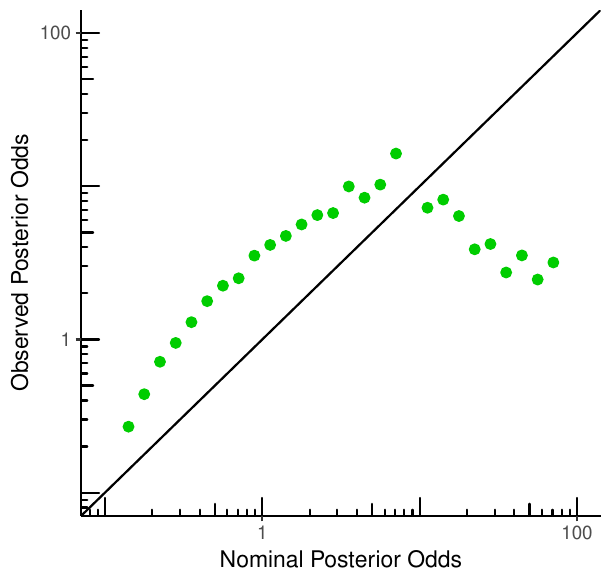}
        \caption{}
        \label{fig:TTestNonSubj b}
    \end{subfigure}
    \caption{Calibration in the $t$-test experiment with fixed values for the means of $\H_0$ and $\H_1$ (Section~\ref{sec:Bayesian t test with default priors}, from $40,000$ replicate experiments). (a) The observed posterior odds as a function of the nominal posterior odds. (b) The observed posterior odds as a function of the nominal posterior odds with optional stopping.}\label{fig:plaatjes TTestNonSubj}
\end{figure}

We generate $10$ observations under both
models. The mean length of the wheat is again set to be $1$ meter with the old
fertilizer, and now we pick a particular value for the mean length of the wheat with the new fertilizer: $130$ centimeters. 
For the variance, we again pick $\sigma^2 =1$. We continue to follow
Rouder's experiment and set our prior odds on $\H_0$ and $\H_1$,
before observing the data, to $1$-to-$1$. We sample $20,000$ replicate
experiments with $10+10$ observations each, $10$ from one of the
hypotheses (normal with mean $1$ for $\H_0$) and $10$ from the other
(normal with mean $\mu=1.3$ for $\H_1$), and we calculate the Bayes
factors. In Figure~\ref{fig:TTestNonSubj a} we see that calibration
is, to some extent, violated: the points follow a line that is still
approximately, but now not precisely, a straight line. Now what
happens in this experiment under optional stopping? We repeated the
experiment with the difference that we sampled more data points until
the posterior odds were at least $10$-to-$1$, or the maximum number of
$25$ data points was reached. In Figure~\ref{fig:TTestNonSubj b} we see
the results:
calibration is now violated significantly --- when we stop early the
nominal posterior odds (on which our stopping rule was based) are
on average significantly higher than the actual, observed posterior odds. We
repeated the experiment with various choices of $\mu$'s within $\H_1$, invariably
getting similar results.\footnote{Invariably, strong calibration is
  violated both with and without optional stopping. In the experiments
  without optional stopping, the points still lie on an increasing and
  (approximately) straight line; the extent to which strong
  calibration is violated --- the slope of the straight line ---
  depends on the effect size. In the experiments with optional
  stopping, strong calibration is violated more strongly in the sense
  that the points do not follow a straight line anymore.}
In mathematical terms, this illustrates that when the stopping time $\tau$ is determined by optional stopping, then, for many $a$ and $\mu'$, 
\begin{equation}\label{eq:calibrated}
\postodds{\mu = \mu', \text{``$\postodds{x_1,\ldots, x_{\tau}} = a$''}} \ \text{is very different from\ }\   a,
\end{equation}
We conclude
that strong calibration for the parameter of interest $\mu$ is
violated somewhat for fixed sample sizes, but much more strongly under
optional stopping. We did similar experiments for a different model
with discrete data (see the Appendix),
once again getting the same result. We
also did experiments in which the means of $\H_1$ were sampled from a
different prior than the Cauchy: this also yielded plots which showed
violation of calibration. Our experiments are all based on a one-sample $t$-test; experiments with a two-sample $t$-test
and ANOVA (also with the same overall mean for both $\H_0$ and $\H_1$)
yielded severe violation of strong calibration under optional stopping
as well.

\paragraph{The Issue}
Why is this important? When checking Rouder's prior-based calibration,
we sampled the effect size from a Cauchy distribution, and then we
sampled data from the realized effect size. We repeated this procedure
many times to approximate the distribution on posterior
odds by a histogram analogous to that in
Figure~\ref{fig:Rouder a}.  But do we really believe that such a
histogram, based on the Cauchy prior, accurately reflects our beliefs
about the data?  The Cauchy prior was advocated by Jeffreys for the
effect size corresponding to a location parameter $\mu$ because it has
some desirable properties in hypothesis testing, i.e.\ when comparing
two models \citep{Ly2016}. For estimating a one-dimensional location
parameter directly, Jeffreys (like most objective Bayesians) would
advocate an improper uniform prior on $\mu$. Thus, objective Bayesians
may {\em change their prior depending on the inference task of
  interest}, even when they are dealing with data representing the
same underlying phenomenon. It does then not seem realistic to study
what happens if data are sampled from the prior; {\em the prior is
  used as a {\em tool\/} in inferring likely parameters or hypotheses,
  and not to be thought of as something that prescribes how actual
  data will arise or tend to look like}.  This is the first reason why
it is interesting to study not just prior calibration, but also strong
calibration for the parameter of interest. One might object that the
sampling from the prior done by Rouder, and us, was only done to
illustrate the mathematical expression (\ref{eq:calibrateb}); perhaps
sampling from the prior is not realistic but (\ref{eq:calibrateb}) is
still meaningful? We think that, because of the mathematical
equivalence, it does show that the relevance of (\ref{eq:calibrateb})
is questionable as soon as we use default priors.

Prior calibration in terms of (\ref{eq:calibrateb}) --- which indeed still holds\footnote{Note though that strong calibration still fails.} ---  {\em would\/}
be meaningful if a Cauchy prior really
described our prior beliefs about the data in the subjective Bayesian
sense (explained in Section~\ref{sec:other}).
But in this particular setup, the Cauchy distribution is highly unrealistic: it is a heavy tailed distribution, which means that the
probability of getting very large values is not negligible, and it is
very much higher than with, say, a Gaussian distribution. To make the
intuition behind this concrete, say that we are interested in
measuring the height of a type of corn that with the old fertilizer
reaches on average $2$ meters. The probability that a new fertilizer
would have a mean effect of $6$ meters or more under a standard Cauchy
distribution would be somewhat larger than one in twenty. For
comparison: under a standard Gaussian, this is
as small as $9.87 \cdot 10^{-10}$. Do we really believe that it is quite
probable (more than one in twenty) that the fertilizer will enable the
corn to grow to $8$ meters \emph{on average}? Of course we could use a Cauchy with a different spread, but which one? 
Default Bayesians have emphasized that such choices should be made subjectively (i.e.\ based on informed prior guesses), but whatever value one choices, the chosen functional form of the prior (a Cauchy has, e.g., no variance) severely restricts the options, making any actual choice to some extent arbitrary.
While growing crops (although a standard example in this context) may
be particularly ill-suited to be modeled by heavy-tailed
distributions, the same issue will arise with many other possible
applications for the default Bayesian $t$-test: one will be practically
sure that the effect size will not exceed certain values (not too large,
not too small, certainly not negative), but it may be very hard to
specify exactly which values. As a purely objective Bayesian,
this need not be such a big problem - one resorts to the default prior
and uses it anyway; but one has to be aware that in that case,
sampling from the prior --- as done by Rouder --- is not
meaningful anymore, since the data one may get may be quite atypical
for the underlying process one is modeling. 

In practice, most Bayesians are pragmatic, striking a balance between
`flat', `uninformative' priors, prior knowledge and ease of
computation. In the present example, they might put a Gaussian prior
with mean $\mu$ on the effect size instead, truncated at $0$ to
avoid negative means. But then there is the question what variance this
Gaussian should have --- as a pragmatic Bayesian, one has to
acknowledge that there will always be arbitrary or `convenience'
aspects about one's priors. This is the second reason why it is
interesting to study not just prior calibration, but also strong
calibration for the parameter of interest.

Thus, both from a purely objective and from a pragmatic Bayesian point
of view, strong calibration is important. Except for nuisance
parameters with Type~0 priors, we cannot expect it to hold precisely
(see \cite{GuHM16} for a related point) --- but this is fine; like
with any sensitivity or robustness test, we acknowledge that our prior
is imperfect and we merely ask that our procedure remains reasonable,
not perfect. And we see that by and large this is the case if we use a
fixed sample size, but not if we perform optional stopping. In our
view this indicates that for pragmatic Bayesians using default priors,
there is a real problem with optional stopping after all. However,
within the taxonomy defined above, we implicitly used Type I priors
(Cauchy) here. Default priors are often of Type II, and then,
as we will see, the problems
get significantly worse.

As a final note, we note that in our strong calibration experiment, we
chose parameter values here which we deemed `reasonable', by this we
mean values which reside in a region of large prior density ---
i.e.\ we sampled from $\mu$ that are not too far from $\mu_0$. Sampling
from $\mu$ in the tails of the prior would be akin to `really
disbelieving our own prior', and would be asking for trouble. We
repeated the experiment for many other values of $\mu$ not too far
from $\mu_0$ and always obtained similar results. Whether our choices
of $\mu$ are truly reasonable is of course up to debate, but we feel
that the burden of proof that our values are `unreasonable' lies with
those who want to show that Bayesian methods can deal with optional
stopping even with default priors.

\subsection{Example 3: Bayesian linear regression and Type II Priors}\label{sec:regression examples}
We further extend the previous example to a setting of linear regression with fixed design. We  employ 
the default Bayes factor for regression from the $\texttt{R}$ package $\texttt{Bayesfactor}$ \citep{BayesFactor}, based on \citet{Liang2008} and \citet{Zellner1980}, see also \citet{RouderRegression2012}. This function uses as default prior Jeffreys' prior for the intercept $\mu$ and the variance ($\pjef(\mu, \sigma) \sim 1 / \sigma$), and a mixture of a normal and an inverse-gamma distribution for the regression coefficients, henceforth \emph{g-prior}:
\begin{align} \nonumber
y &\sim \text{N}\left(\mu + X\beta , \sigma^2 \right), \\ \label{eq:regression: prior on beta}
\beta &\sim \text{N}\left(0, g \sigma^2 n(X'X)^{-1} \right), \\ \nonumber
g &\sim \text{IG}\left( \frac{1}{2}, \frac{\sqrt{2}}{8} \right).
\end{align}
Since the publication of \citet{Liang2008}, this prior has become very
popular as a default prior in Bayesian linear regression. Again we
provide an example concerning the growth of wheat.  Suppose a
researcher wants to investigate the relationship between the level of
a fertilizer, and the growth of the crop. We can model this experiment by linear regression with fixed design. We add different levels of the fertilizer to pots with seeds: the first pot gets a dose of $0.1$, the second $0.2$, ans so on up to the level $2$. These are the $x$-values (covariates) of our simulation experiment. If we would like to repeat the examples of the previous sections and construct the calibration plots, we can generate the $y$-values --- the increase or decrease in length of the wheat from the intercept $\mu$ --- according to the proposed priors in Eq.~(\ref{eq:regression: prior on beta}). First we draw a $g$ from an inverse gamma distribution, then we draw a $\beta$ from the normal prior that we construct with the knowledge of the $x$-values, and we compute each $y_i$ as the product of $\beta$ and $x_i$ plus Gaussian noise. 

\begin{figure}
  \centering
    \begin{subfigure}[b]{0.4\textwidth}
 \includegraphics[width=\textwidth]{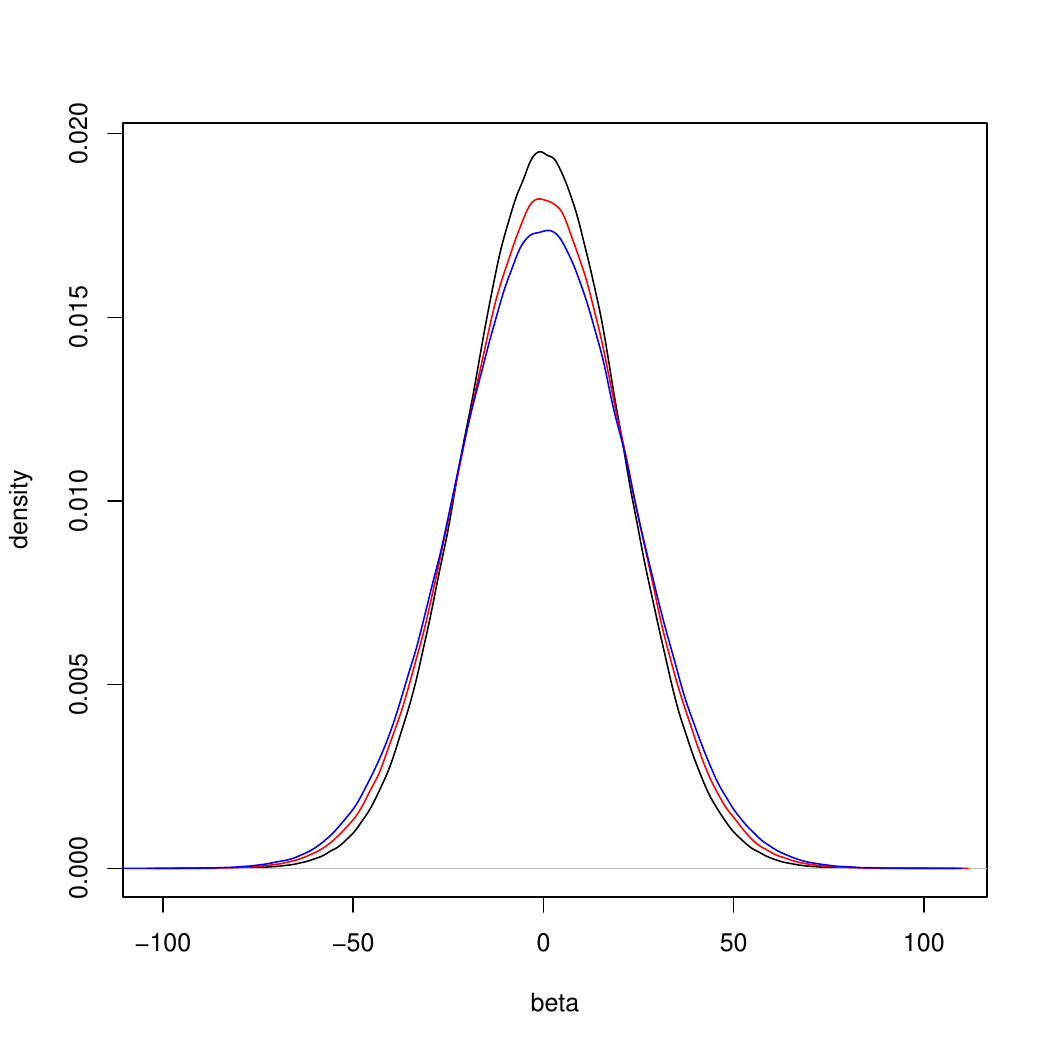}
        \caption{}
        \label{fig:Gpriors}
    \end{subfigure}
    \hfill
    \begin{subfigure}[b]{0.4\textwidth}
        \includegraphics[width=\textwidth]{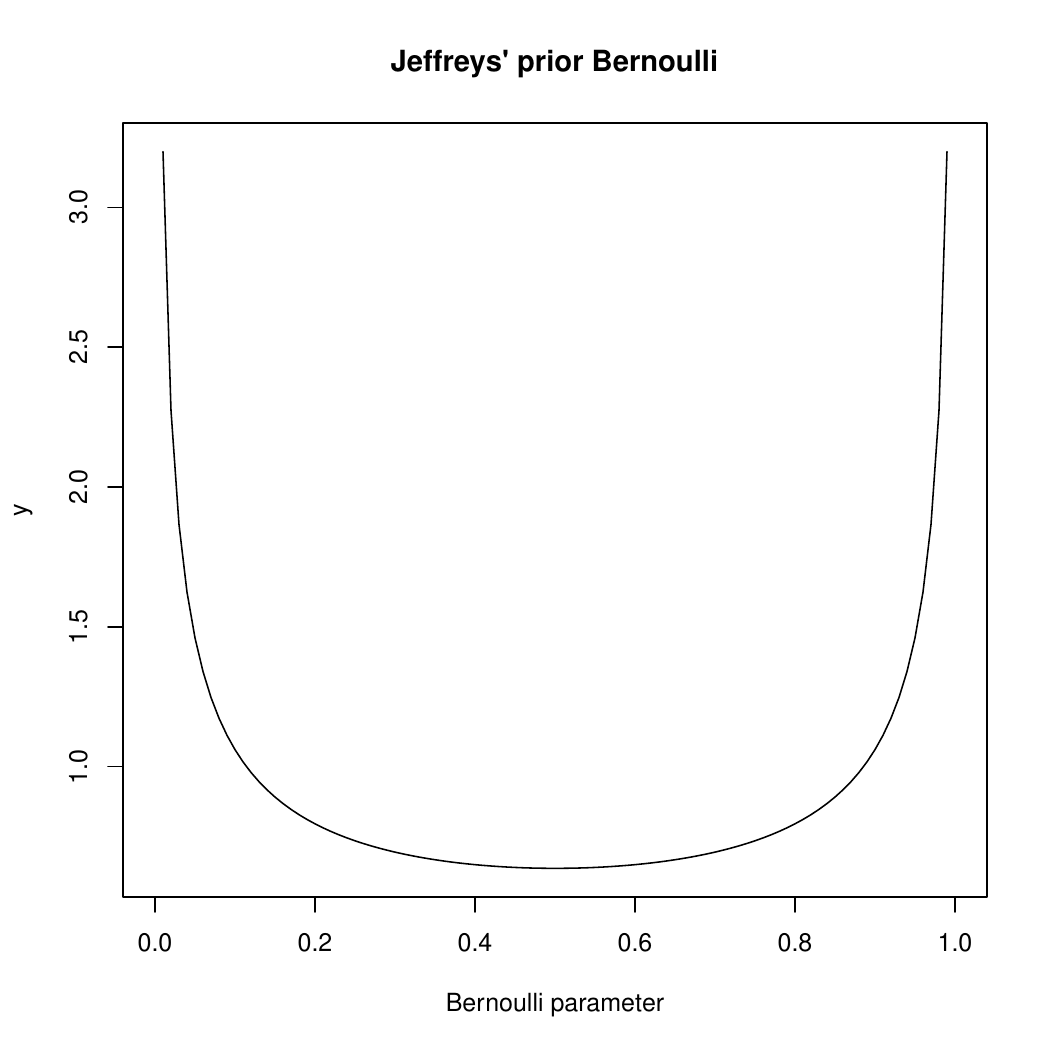}
        \caption{}
        \label{fig:Bernoulli Jeffreys Prior}
    \end{subfigure}
    \caption{Default priors that depend on aspects of the experimental
      setup: (a) g-priors for the regression example of
      Section~\ref{sec:regression examples} with different sample
      sizes: $n=20$ (black), $n=23$ (red) and $n=34$ (blue). (b)
      Jeffreys' prior for the Bernoulli model for the specific case
      that $n$ is fixed in advance (no optional
      stopping): a Beta $(1/2,1/2)$ distribution.}\label{fig:Gpriors en Bernoulli}
\end{figure}

As we can see in Equation~\ref{eq:regression: prior on beta}, the
prior on $\beta$ contains a scaling factor that depends on the
experimental set-up --- while it does not directly depend on the
observations ($y$-values), it does depend on the design/covariates
($x$-values). If there is no optional stopping, then for a pragmatic
Bayesian, the dependency on the $x$-values of the data is convenient
to achieve appropriate scaling; it poses no real problems, since the
whole model is conditional on $X$: the levels of fertilizer we
administered to the plants. But under optional stopping, the
dependency on $X$ does become problematic, {\em for it is unclear
  which prior she should use!} If initially a design with $40$ pots
was planned (after each dose from $0.1$ up to $2$, another row of
pots, one for each dose is added), but after adding three pots to the
original twenty (so now we have two pots with the doses $0.1, 0.2$ and
$0.3$, and one with each other dose), the researcher decides to check
whether the results already are interesting enough to stop, should she
base her decision on the posterior reached with prior based the
initially planned design with $40$ pots, or the design at the moment
of optional stopping with $23$ pots? This is not clear, and it does
make a difference, since the g-prior changes as more $x$-values become
available. In Figure~\ref{fig:Gpriors} we see three g-priors on the
regression coefficient $\beta$ for the same fixed value of $g$, the
same $x$-values as described in the fertilizer experiment above, but
increasing sample size. First, each dose is administered to one plant,
yielding the black prior distribution for $\beta$. Next, $3$ plants
are added to the experiment, with doses $0.1, 0.2$ and $0.3$, yielding
the red distribution: wider and less peeked, and lastly, another $11$
plants are added to the experiment, yielding the blue distribution
which puts even less prior mass close to zero.

This problem may perhaps be pragmatically `solved' in practice in two
ways: either one could, as a rule, base the decision to stop at sample
size $n$ always using the prior for the given design at sample size
$n$; or one could, as a rule, always use the design for the maximum
sample size available. It is very unclear though whether there is any
sense in which any of these two (or other) solutions `handle optional
stopping' convincingly. In the first case, the notion of prior
calibration is ill-defined, since $\postodds{x_1,\ldots, x_{\tau}}$ in
(\ref{eq:calibrateb}) is ill-defined (if one tried to illustrate
(\ref{eq:calibrateb}) by sampling, the procedure would be undefined
since one would not know what prior to sample from until after one has
stopped); in the second, one can perform it (by sampling $\beta$ from
the prior based on the design at the maximum sample size), but it
seems rather meaningless, for if, for some reason or other, even more
data were to become available later on, this would imply that the
earlier sampled data were somehow `wrong' and would have to be
replaced.

What, then, about strong calibration?
Fixing particular,
`reasonable' values of $\beta$ is meaningful in this regression
example. However (figures omitted), when we pick reasonable values for
$\beta$ instead of sampling $\beta$ from the prior, we obtain again
the conclusion that strong calibration is, on one hand, violated
significantly under optional stopping (where the prior used in the
decision to stop can be defined in either of the two ways defined
above); but on the other hand, only violated mildly for fixed sample
size settings.
Using the taxonomy above, we conclude that optional stopping is a significant problem for Bayesians with Type-II priors.

\subsection{Discrete Data and Type-II Priors} 
\label{sec:bernoulli}
Now let us turn to discrete data: 
we test whether a coin is fair or not.  
The data $D$ consist of a sequence of $n_1$ ones and $n_0$ zeros. 
Under ${\cal H}_0$, the data are i.i.d.\ Bernoulli$(1/2)$; under ${\cal H}_1$
they can be Bernoulli$(\theta)$ for any $ 0 \leq \theta \leq 1$ except $1/2$, $\theta$ representing the bias of the coin. One standard objective and default Bayes method (in this case coinciding with an {\em MDL (Minimum Description Length) method}, \citep{Grunwald2007}) is to use Jeffreys' prior for the Bernoulli model within ${\cal H}_1$. For fixed sample sizes, this prior is proper, and is
given by
\begin{equation}\label{eq:bernoulli}
\pjef(\theta) = \frac{1}{\sqrt{\theta(1- \theta)}} \cdot \frac{1}{\pi}, 
\end{equation}
where the factor $1/\pi$ is for normalization; see
Figure~\ref{fig:Bernoulli Jeffreys Prior}. If we repeat Rouder's
experiment, and sample from this prior, then the probability that we
would pick an extreme $\theta$, within $0.01$ of either $1$ or $0$,
would be about $10$ times as large as the probability that we would
pick a $\theta$ within the equally wide interval $[0.49,0.51]$. But,
lacking real prior knowledge, do we really believe that such extreme
values are much more probable than values around the middle? Most
people would say we do not: under the subjective interpretation, i.e.\
if one really believes one's prior in the common interpretation of
`belief' given in Section~\ref{sec:other}, then such a prior would
imply a willingness to bet at certain stakes. Jeffreys' prior is
chosen in this case because it has desirable properties such as
invariance under reparameterization and good frequentist properties,
but not because it expresses any `real' prior belief about some
parameter values being more likely than others. This is reflected in
the fact that in general, it depends on the stopping rule. Using the
general definition of Jeffreys' prior (see e.g.\ \citet{Berger85}), we
see, for example, that in the Bernoulli model, if the sample size is
not fixed in advance but depends on the data (for example, we stop
sampling as soon as three consecutive $1$s are observed), then, as a
simple calculation shows, Jeffreys' prior changes and even becomes
improper \citep{Jordan10}.

In the Appendix we give another
example of a common discrete setting, namely the $2\times2$ contingency
table. Here the null hypothesis is a Bernoulli model and its parameter
$\theta$ is intuitively a nuisance parameter, and thus strong
calibration relative to this parameter would be especially
desirable. However, the Bernoulli model does not admit a group
structure, and hence neither Jeffreys' nor any other prior we know of can serve
as a Type~0 prior, and strong calibration can presumably not be
attained --- the experiments show that it is certainly not attained if
the default Gunel and Dickey Bayes factors \citep{Jamil2016} are used
(these are Type-II priors, so we need to be careful about what prior
to use in the strong calibration experiment; see
the Appendix for details).

\section{Other Conceptualizations of Optional Stopping}
\label{sec:other}
We have seen several problems with optional stopping under default and
pragmatic priors. Yet it is known from the literature
that, in some senses, optional stopping is indeed no problem for
Bayesians \citep{Lindley1957,Savage1972,Edwards1963,Good1991}. What
then, is shown in those papers? Interestingly, different authors show
different things; we consider them in turn. 

\subsection{Subjective Bayes optional stopping}
The Bayesian pioneers \citet{Lindley1957} and \citet{Savage1972}
consider a purely subjective Bayesian setting, appropriate if one
truly believes one's prior (and at first sight completely disconnected
from strong calibration --- but see the two quotations further below). But what does
this mean?  According to De Finetti, one of the two main founding
fathers of modern, subjective Bayesian statistics, this implies a
willingness to bet at small stakes, at the odds given by the
prior.\footnote{Savage, the other father, employs a slightly different
  conceptualization in terms of preference orderings over outcomes,
  but that need not concern us here.}  For example, a subjective
Bayesian who would adopt Jeffreys' prior $\pjef$ for the Bernoulli
model as given by (\ref{eq:bernoulli}) would be willing to accept a
gamble that pays off when the actual parameter lies close to the
boundary, since the corresponding region has substantially higher
probability, cf. the discussion underneath
Eq.~(\ref{eq:bernoulli}). For example, a gamble where one wins $11$
cents if the actual Bernoulli parameter is in the set $[0,0.01] \cup
[0.99,1]$ and pays $100$ cents if it is in the set $[0.49,0.51]$ and
neither pays nor gains otherwise would be considered
acceptable\footnote{One might object that actual Bernoulli parameters
  are never revealed and arguably do not exist; but one could replace
  the gamble by the following essentially equivalent gamble: a
  possibly biased coin is tossed $10,000$ times, but rather than the
  full data only the average number of $1s$ will be revealed. If it is
  in the set $[0,0.01] \cup [0.99,1]$ one gains 11 cents and if it is
  in the set $[0.49,0.51]$ one pays $100$ cents. If one really
  believes Jeffreys' prior, this gamble would be considered
  acceptable.}  because this gamble has positive expected gain under
$\pjef$.  We asked several Bayesians who are willing to use Jeffreys'
prior for testing whether they would also be willing to accept such a
gamble; most said no, indicating that they do not interpret Jeffreys
prior the way a subjective Bayesian would.\footnote{Another example is
  the Cauchy prior with scale one on the standardized effect size
  \citep{Rouder2012}, as most would agree that this is not realistic
  in psychological research. Thanks to an anonymous reviewer for
  pointing this out.} 

Now, if one adopts priors one really believes in in the above gambling
sense, then Bayesian updating from prior to
posterior is not affected by the employed stopping rule \citep{OptionalStoppingTechnical}; one ends up
with the same posterior if one had decided the sample size $n$ in
advance or if it had been determined, for example, because one was
satisfied with the results at this $n$. In this sense a subjective
Bayesian procedure does not depend on the stopping rule (as we
have seen, this is certainly not the case in general for default Bayes
procedures). This is the main point concerning optional stopping of
\citet{Lindley1957}, also made by e.g.\
\citet{Savage1972,BernardoS94}, among many others. A second point made
by \citet[p.\ 192]{Lindley1957} is that the decisions a Bayesian makes
will ``not, {\em on average}, be in error, when ignoring the stopping
rule''. Here the ``average'' is really an expectation obtained by
integrating $\theta$ over the prior, and then the data $D$ over the distribution $\pr(D \mid {\theta})$, making
this claim very similar to prior calibration (\ref{eq:calibrateb})  --- once again, the claim is
correct, but works only if one believes that
sampling (or taking averages over) the prior gives rise to data of the type one would really expect; and if one would not be willing to bet based on the prior in the above sense, it indicates that perhaps one doesn't really expect that data after all.

We cannot resist to add here that, while for a subjective Bayesian,
prior-based calibration is
sensible, even the founding fathers of subjective Bayes gave a warning
against taking such a prior too seriously:\footnote{Many thanks to
  Chris Holmes for bringing these quotations to our attention.}
\begin{quote}
`` Subjectivists should feel obligated to recognize that any opinion (so
much more the initial one) is only vaguely acceptable... So it is
important not only to know the exact answer for an exactly specified
initial problem, {\em but what happens changing in a reasonable
neighborhood the assumed initial opinion}'' De Finetti, as quoted
by \citet{Dempster1975}. --- note that when we checked for strong calibration, we took parameter values $\mu$ which were not too unlikely under the prior, which one may perhaps view as `a reasonable neighborhood of the initial opinion'.  

`` ...in practice the theory of personal probability is supposed to be an
idealization of one's own standard of behavior; the idealization
is often imperfect in such a way that an aura of vagueness is
attached to many judgments of personal probability...''  \citep{Savage1972}.
\end{quote}

Hence, one would expect that even a subjectivist would be interested
in seeing what happens under a sensitivity analysis, for example
checking for strong rather than prior-based calibration of the
posterior. And even a subjectivist cannot escape the conclusion from
our experiments that optional stopping leads to more brittle (more
sensitive to the prior choice) inference than stopping at a fixed $n$. 

\subsection{Frequentist optional stopping under $\H_0$}\label{sec:frequentist optional stopping}
Interestingly, some other well-known Bayesian arguments claiming that
`optional stopping is no problem for Bayesians' really show that some
Bayesian procedures can deal, in some cases, with optional stopping in
a different, frequentist sense. These include
\citet{Edwards1963,Good1991} and many others (the difference between
this justification and the above one by \citet{Lindley1957} roughly
corresponds to Example 1 vs.\ Example 2 in the appendix to
\citep{Wagenmakers2007}). We now explain this frequentist notion of optional
stopping, emphasizing that some (but --- contrary to what is claimed
--- by no means all!) tests advocated by Bayesians {\em do\/} handle
optional stopping in this frequentist sense.
%

The (or at least, `a common') frequentist interpretation of handling
optional stopping is about controlling the Type I error of an
experiment. A Type I error occurs when we reject the null hypothesis
when it is true, also called a \emph{false positive}. The probability
of a Type I error for a certain test is called the \emph{significance
  level}, usually denoted by $\alpha$, and in psychology the value of
$\alpha$ is usually set to $0.05$. A typical classical hypothesis test
computes a test statistic from the data and uses it to calculate a
p-value. It rejects the null hypothesis if the p-value is below the
desired Type I error level $\alpha$. For other types of hypothesis
tests, it is also a crucial property to control the Type I error, by
which we mean that we can make sure that the probability of making a
Type I error remains below our chosen significance level $\alpha$. The
frequentist interpretation of handling optional stopping is that the
Type I error guarantee holds if we do not determine the sampling plan
--- and thus the stopping rule --- in advance, but we may stop when we
see a significant result. As we know, see e.g.\
\citet{Wagenmakers2007}, maintaining this guarantee under optional
stopping is not possible with most classical p-value based hypothesis
tests. 

At first sight none of this seems applicable to Bayesian tests, which output posterior odds rather than a p-value. However, in the case that ${\cal H}_0$ is {\em simple\/} (containing just one hypothesis, as in Example 0), there is a well-known intriguing connection between Bayes factors and Type I error probabilities: 
--- if we reject $H_0$ iff the posterior odds in
favor of $\H_0$ are smaller than some fixed $\alpha$, then we are guaranteed a Type I error of at most
$\alpha$. And interestingly, this holds not just for fixed sample sizes but even under optional stopping. Thus, if one adopts the rejection rule above (reject iff the posterior odds are smaller than a fixed $\alpha$),  {\em for {\em simple\/} ${\cal H}_0$,
frequentist optional stopping is no problem for
Bayesians}. This is what was noted by
\citet{Edwards1963} (using a different terminology) and
\citet{Good1991}, based on what \citet{Sanborn2014} call the
\emph{universal bound}, and what in probability theory is known as
\emph{Doob's maximal inequality\/} \citep{Doob1971}; see also  \citet{Vovk2011} and
\citet{VdPas2016}.

But what happens if $\H_0$ is composite?
As was only shown very recently \citep{OptionalStoppingTechnical}, the
Bayes factor still handles optional stopping in the frequentist sense
if {\em all\/} free parameters in $\H_0$ are nuisance parameters
observing a group structure and equipped with the corresponding Type~0
prior and are shared with $\H_1$, an example being Jeffreys' Bayesian
$t$-test of Section~\ref{sec:Bayesian t test with default priors}. As
explained by \citet{OptionalStoppingTechnical}, for general priors and
composite ${\cal H}_0$ though, this is typically not the case; for
example, the Gunel-Dickey default Bayes factors for$2\times2$ tables
(another composite $\H_0$) cannot handle  optional
stopping in the frequentist sense.


\paragraph{An Empirical Frequentist Study of Bayesian Optional
  Stopping}
\citet{Schoenbrodt2017} performed a thorough simulation study to
analyze frequentist performance of optional stopping with Bayes
factors both under $\H_0$ and under $\H_1$. They confined their
analysis to the Bayesian $t$-test, i.e.\ our Example 2, and found
excellent results for the Bayesian optional stopping procedure {\em
  under a certain frequentist interpretation\/} of the Bayes factors (posterior
  odds). As to optional stopping under $\H_0$ (concerning Type I
error), this should not surprise us: in the Bayesian $t$-test,
all free parameters in $\H_0$ are equipped with Type~0 priors, which,
as we just stated, can handle optional stopping. We thus feel that one
should be careful in extrapolating their results to other models such
as those for contingency tables, which do not admit such priors.  As
to optional stopping under $\H_1$, the authors provide a table showing
how, for any given effect size $\delta$ and desired level of Type II
error $\beta$, a threshold $B$ can be determined such that the standard Bayesian $t$-test with
(essentially) the following optional stopping and decision rule, has Type II error $\beta$:
\begin{quote}
Take at least 20 data points. After that stop as soon as  posterior odds are larger than $B$ or smaller than
$1/B$; accept $\H_0$ if they are smaller than $1/B$, and reject $\H_0$ if
larger than $B$.  
\end{quote}
For example, if $\delta \geq 0.3$ and one takes $B=7$ then the Type II
error will be smaller than $4\%$ (see their Table 1). They also
determined the average sample size needed before this procedure stops, and
noted that this is considerably smaller than with the standard
$t$-test optimized for the given desired levels of Type I and Type II
error and a priori expected effect size. Thus, if one determines the
optional stopping threshold $B$ in the Bayesian $t$-test based on
their table, one can use this Bayesian procedure as a frequentist
testing method that significantly improves on the standard $t$-test in
terms of sample size. Under {\em this\/} frequentist interpretation
(which relies on the specifics of a table), optional stopping with the
$t$-test is indeed unproblematic.  Note that this does not contradict
our findings in any way: our simulations show that if, when sampling,
we fix an effect size in $\H_1$, then the posterior is biased under
optional stopping, which means that we cannot interpret the posterior
in a {\em Bayesian\/} way.

\section{Discussion and Conclusion}\label{sec:discussion, conclusion}

When a researcher using Bayes factors for hypothesis testing truly
believes in her prior, she can deal with optional stopping in the
Bayesian senses just explained.  However, these senses become
problematic for every test that makes use of default priors, including
all default Bayes factor tests advocated within the Bayesian
Psychology community. Such `default' or `objective' priors cannot be
interpreted in terms of willingness to bet, and sometimes (Type II
priors) depend on aspects of the problem at hand such as the stopping
rule or the inference task of interest. To make sense of such priors
generally, it thus seems necessary to {\em restrict\/} their use to
their appropriate domain of reference --- for example, Jeffreys' prior
for the Bernoulli model as given by (\ref{eq:bernoulli}) is okay for
Bayes factor hypothesis testing with fixed sample size, but not for
more complicated stopping rules. This idea, which is unfortunately
almost totally lacking from the modern Bayesian literature, is the
basis of a novel theory of the very concept of probability called {\em
  Safe Probability\/} which is being developed by one of us
\citep{Grunwald13,Grunwald17}. That (mis)use of optional stopping is a
serious problem in practice, is shown by, among others,
\citet{john2012measuring}; however, that paper is (implicitly) mostly
about frequentist methods. It would be interesting to investigate to
what extent optional stopping when combined with default Bayesian
methods is actually a problem not just in theory but also in
practice. This would, however, require substantial further study and simulation.

\citet{Rouder2014} argues in response to \citet{Sanborn2014} that the
latter `evaluate and interpret Bayesian statistics as if they were
frequentist statistics', and that `the more germane question is
whether Bayesian statistics are interpretable as \emph{Bayesian
  statistics}'. Given the betting interpretation above, the essence
here is that we need to make a distinction between the purely
subjective and the pragmatic approach: we can certainly not evaluate
and interpret \emph{all} Bayesian statistics as \emph{purely
  subjective} Bayesian statistics, what \citet{Rouder2014} seems to
imply. He advises Bayesians to use optional stopping --- without any
remark or restriction to purely subjective Bayesians, and for a
readership of experimental psychologists who are in general not
familiar with the different flavors of Bayesianism --- as he writes
further on: `Bayesians should consider optional stopping in
practice. [...] Such an approach strikes me as justifiable and
reasonable, perhaps with the caveat that such protocols be made
explicit before data collection'. The crucial point here is that this
can indeed be done when one works with a purely subjective Bayesian
method, but not with the \emph{default Bayes factors} developed for
practical use in social science: both strong calibration and the
frequentist Type I-error guarantees will typically be violated, and
for Bayes factors involving Type II-priors, both prior and strong
calibration are even undefined. In Table~\ref{tab:overview kinds of OS
  with common Default BF} we provide researchers with a simplified
overview of four common default Bayes factors indicating which forms
of optional stopping they can handle.

While some find the purely subjective Bayesian framework unsuitable
for scientific research (see e.g.\ \citet{Berger2006}), others deem it
the only coherent approach to learning from data per se. We do not
want to enter this discussion, and we do not have to, since in
practice, nowadays most Bayesian statisticians tend to use priors
which have both `default' and `subjective' aspects. Basically, one
uses mathematically convenient priors (which one does not really
believe, so they are not purely subjective --- and hence, prior
calibration is of limited relevance), but they are also chosen to be
not overly unrealistic or to match, to some extent, prior knowledge
one might have about a problem.  This position is almost inevitable in
Bayesian practice (especially since we would not like to burden
practitioners with all the subtleties regarding objective and
subjective Bayes), and we have no objections to it --- but it does
imply that, just like frequentists, Bayesians should be careful with
optional stopping. For researchers who like to engage in optional
stopping but care about frequentist concepts such as Type I error and
power, we recommend the {\em safe tests\/} of \cite{GrunwaldHK19}
based on the novel concept of {\em $E$-variables\/}: $E$-variables are
related to, and sometimes coincide with, default Bayes factors, but
tests based on $E$-variables invariably handle a variation of
frequentist optional stopping. For example, the three default Bayes
factors that handle frequentist optional stopping in
Table~\ref{tab:overview kinds of OS with common Default BF} are also
$E$-variables, but there exist other $E$-variables for these three settings
that also handle optional stopping but achieve higher frequentist
power; and there also exists an $E$-variable for contingency tables that,
unlike the default Bayes factor, handles frequentist optional
stopping.

\pagebreak
\paragraph{Open Practices Statement} Since all the data involved in this paper was generated by straightforward computer simulations rather than `real-world' experiments, we did not make the data available. No experiments were done, and hence no experiments were
preregistered.
\begin{table}[]
\centering
\begin{tabular}{ll|lll}
\multicolumn{2}{l|}{}                    & Prior Cal.                  & Strong Calibration & Freq. OS                  \\ \specialrule{.2em}{.1em}{.1em}
\multicolumn{2}{l|}{Default Bayes Factors}                    &             &  & \\ \specialrule{.2em}{.1em}{.1em}   
\multirow{2}{*}{} & \multirow{2}{*}{T-test \citep{Rouder2009}} & \multirow{2}{*}{\cmark \, {\color{red} \bf but...} (I)} & \cmark \,  for $\sigma$\ \ (0)  & \multirow{2}{*}{\cmark} \\ 
                   &                     &                     & \xmark \,\, for $\delta$ (effect size)\ \ (I)  &                     \\ \cline{2-5}
\multirow{2}{*}{} & \multirow{2}{*}{ANOVA \citep{Rouder2012}} & \multirow{2}{*}{\cmark \, {\color{red} \bf but...} (I)   } & \cmark \, for $\mu, \sigma$ \ \ (0) & \multirow{2}{*}{\cmark} \\
                   &                     &                     & \xmark \,\, for $\delta$ (effect size)\ \ (I)  &                     \\ \cline{2-5}
	& Regression & \multirow{2}{*}{\xmark \ \ (II)} & \cmark \, for $\mu, \sigma$ \ \ (0) & \multirow{2}{*}{\cmark} \\
                   &   \citep{RouderRegression2012}               &                     & \xmark \,\, for $\beta$ (effects) \ \ \ \ (II) &                     \\ \cline{2-5}
                                        & Contingency Tables        & \multirow{2}{*}{\xmark \ \ (II)}     & \multirow{2}{*}{\xmark} & \multirow{2}{*}{\xmark}     \\
                                       & \citep{Jamil2016}        &                       &                         &                             \\ \specialrule{.2em}{.1em}{.1em}
\multicolumn{2}{l|}{Bayes Factors with proper, fully}                                                             & \multirow{2}{*}{\cmark}     & \multirow{2}{*}{N/A}   & \multirow{2}{*}{N/A}    \\
\multicolumn{2}{l|}{subjective priors \citep{Rouder2014}}                                                             &                         &                        &
\end{tabular}
\caption{Overview of several common default Bayes Factors (from the $\texttt{R}$-package $\texttt{BayesFactor}$ \citep{BayesFactor}), and their robustness against different kinds of optional stopping (proofs can be found in \cite{OptionalStoppingTechnical}). `Prior Cal.' means `prior calibration' and `Freq. OS' means `frequentist optional stopping'. Between parentheses is the type of prior used (0, I or II), in the taxonomy introduced in this paper. The {\color{red} \bf but..} indicates that, formally, prior calibration works for the priors, yet, because we are in the default setting, the Bayes factor is not fully subjective, so 
  prior calibration is not too meaningful --- which is just the main point of this paper.
}
\label{tab:overview kinds of OS with common Default BF}
\end{table}

\DeclareRobustCommand{\VANDER}[3]{#3}

\clearpage
\bibliographystyle{abbrvnat}
\bibliography{mybib}

\clearpage
\appendix

\section{Example 4: An independence test in a 2x2 contingency table}\label{sec:contingency table exp.}
Suppose that a researcher considers two hypotheses: a null hypothesis
$\H_0$ that states that there is no difference in voting preference
(Democrat or Republican) between men and women, and an alternative
hypothesis $\H_1$ stating that men's voting preferences differ from
the women's preferences. Both hypotheses are {composite} --- we may
think of a Bernoulli model for $\H_0$: the data are i.i.d.\ with a
fixed probability of $1$ (voting Democrat). We are however not
interested in the percentage of the persons voting for the
Democrats. We are, instead, only interested to learn if this
percentage is \emph{equal} for men and women or not. Thus our null
hypothesis $\H_0$ consists of all Bernoulli distributions (all
possible biases of the coin, infinitely many between $0$ and $1$)
where the model for the men is the same as for the women. Our
alternative hypothesis is composite as well: all the sets of two
Bernoulli distributions --- one for the men and one for the women ---
that are not equal. Thus, the Bernoulli parameter in $\H_0$ is not a
parameter of interest; instead, at least intuitively, it is a nuisance
parameter similar to the variance in Example 1; however, it does not observe a group structure and a Type 0-prior for this parameter does not exist. \\

Once again we follow Rouder's experiments closely. We now use the
\emph{Default Gunel and Dickey Bayes Factors for Contingency
	Tables}
\citep{Jamil2016}, which employs specific default choices for the
priors within $\H_0$ and $\H_1$, depending on four different sampling
schemes (see Section~\ref{sec:details experiments cont tables} for the details).  We immediately run into a problem similar to the problems
described with the $g$-prior and Jeffreys' prior for Bernoulli: which
prior we should choose depends on the sampling plan itself. Based on
earlier work by \cite{Dickey1974} (GD from now on),
\citet{Jamil2016} provide different default priors depending on
whether the sample size $n$ and/or some of the four counts (number of
men/women voting democratic/republican) are fixed in advance. For the
case that none of these are fixed in advance, they provide a prior
which assumes that the four counts are all Poisson distributed; see
the next section for details. Intuitively, none of these priors seem
to be compatible with the very idea of `optional stopping' and
prior-based calibration under optional stopping cannot be tested
(since it is not clear what prior to sample from --- a Type II-problem
in our earlier terminology).  Still, to check the claim that `optional
stopping is no problem for Bayesians' we will again check whether {\em
	strong\/} calibration holds with and without optional stopping. We
display here the results of an experiment with the prior advocated for
the case in which neither $n$ nor any of the counts are assumed to be
fixed in advance, since this seems the choice least incompatible with
optional stopping. To avoid discussion on this issue though, we also
performed the experiments with the priors advocated for other sampling
schemes and combinations of different sampling schemes, which led to
very similar results.\\

We will again fix some `reasonable' parameter values in each model:
when sampling from $\H_0$, we really sample from $\theta = 1/2$, i.e.\
we suppose that $50\%$ of either gender prefers the Democrats. When we
sample from $\H_1$, we suppose that $45\%$ of the men prefers the
Democrats, but for the women it is as much as $55\%$. If there are
equally many men as women, under both hypotheses the average
percentage is equal. Like Rouder, we set our prior odds to $1$-to-$1$.\\

\begin{figure}
	\centering
	\begin{subfigure}[b]{0.5\textwidth}
		\includegraphics[width=\textwidth]{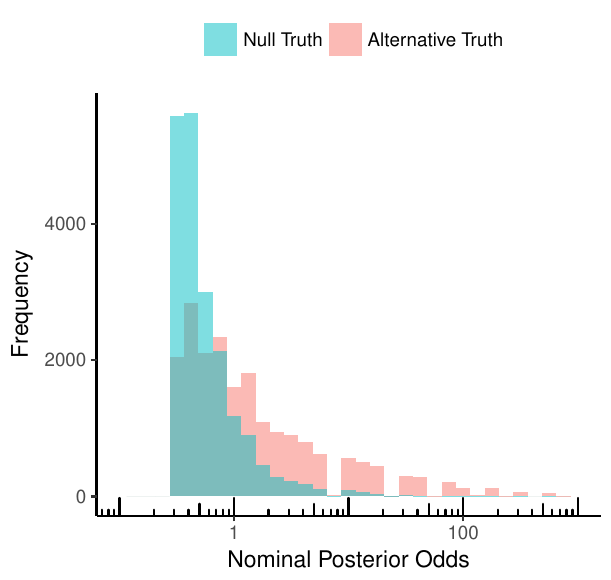}
		\caption{}
		\label{fig:a}
	\end{subfigure}
	\hfill
	\begin{subfigure}[b]{0.4\textwidth}
		\includegraphics[width=\textwidth]{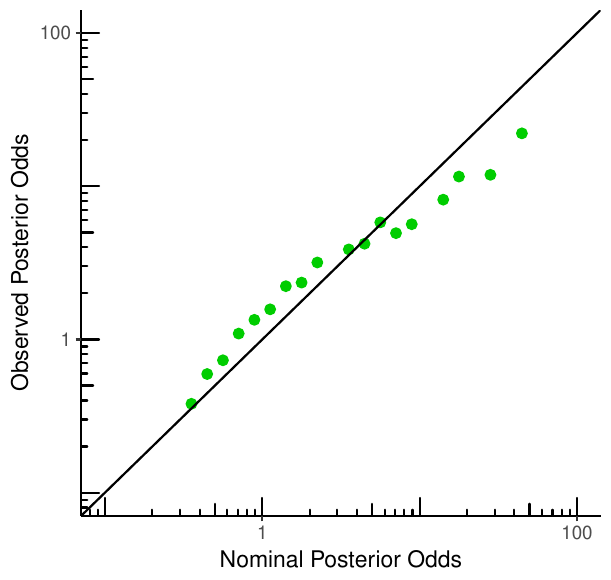}
		\caption{}
		\label{fig:b}
	\end{subfigure}
	\vskip\baselineskip
	\begin{subfigure}[b]{0.5\textwidth}
		\includegraphics[width=\textwidth]{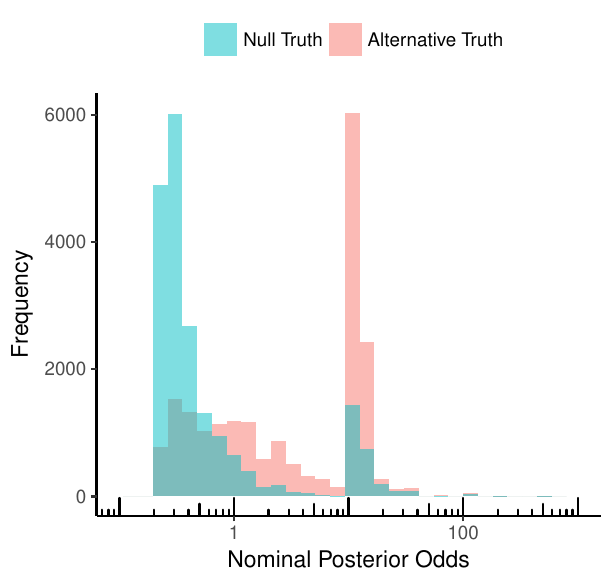}
		\caption{}
		\label{fig:c}
	\end{subfigure}
	\hfill
	\begin{subfigure}[b]{0.4\textwidth}
		\includegraphics[width=\textwidth]{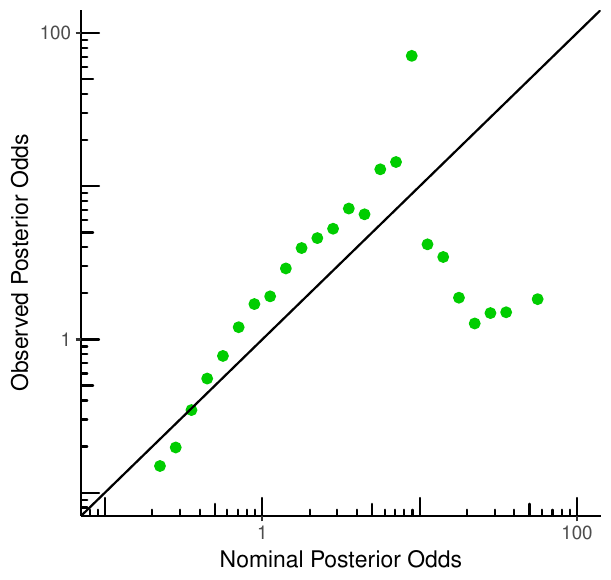}
		\caption{}
		\label{fig:d}
	\end{subfigure}
	\caption{Calibration of the contingency table experiment, Section~\ref{sec:contingency table exp.}, from $20.000$ replicate experiments. (a) The distribution of posterior odds as a histogram under $\H_0$ and $\H_1$. (b) The observed posterior odds as a function of the nominal posterior odds. (c) Distribution of the posterior odds with optional stopping. (d) The observed posterior odds as a function of the nominal posterior odds with optional stopping.}\label{fig:plaatjes}
\end{figure}

We simulate $20.000$ replicate experiments of $100+100$ samples each, from both $\H_0$ and $\H_1$, and we calculate the Bayes Factors. We construct the histograms and the plots with the odds as before. We can check the calibration in Figure~\ref{fig:b}: we can see that the nominal posterior odds agree roughly with the observed posterior odds. In Figure~\ref{fig:d} however, we see the same plot where we did the same experiment with optional stopping. We can clearly see that even the rough linear relationship from Figure~\ref{fig:b} is completely gone. For this example, we can conclude as well that strong calibration is violated.\\

\begin{figure}
	\centering
	\begin{subfigure}[b]{0.4\textwidth}
		\includegraphics[width=\textwidth]{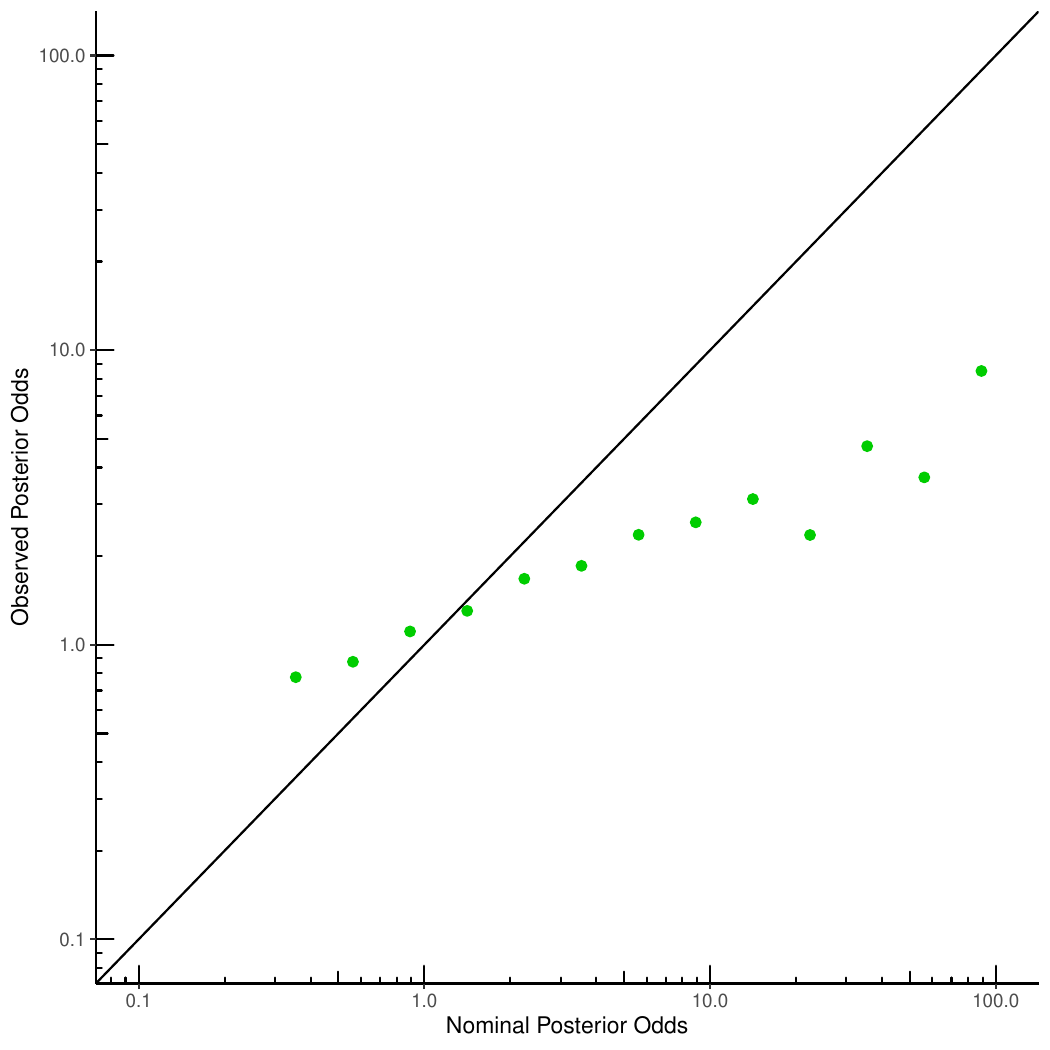}
		\caption{}
		\label{fig:Extra a}
	\end{subfigure}
	\hfill
	\begin{subfigure}[b]{0.4\textwidth}
		\includegraphics[width=\textwidth]{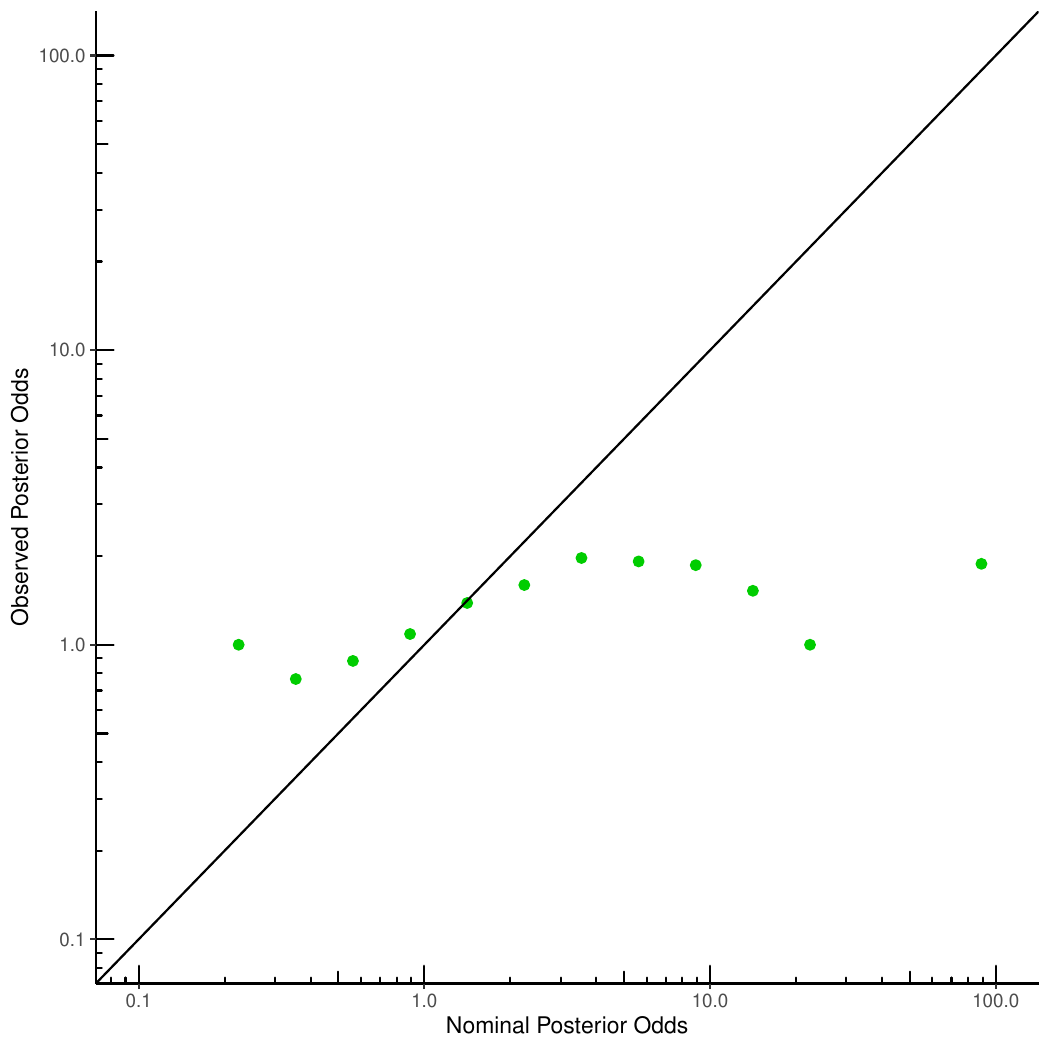}
		\caption{}
		\label{fig:Extra b}
	\end{subfigure}
	\vskip\baselineskip
	\begin{subfigure}[b]{0.4\textwidth}
		\includegraphics[width=\textwidth]{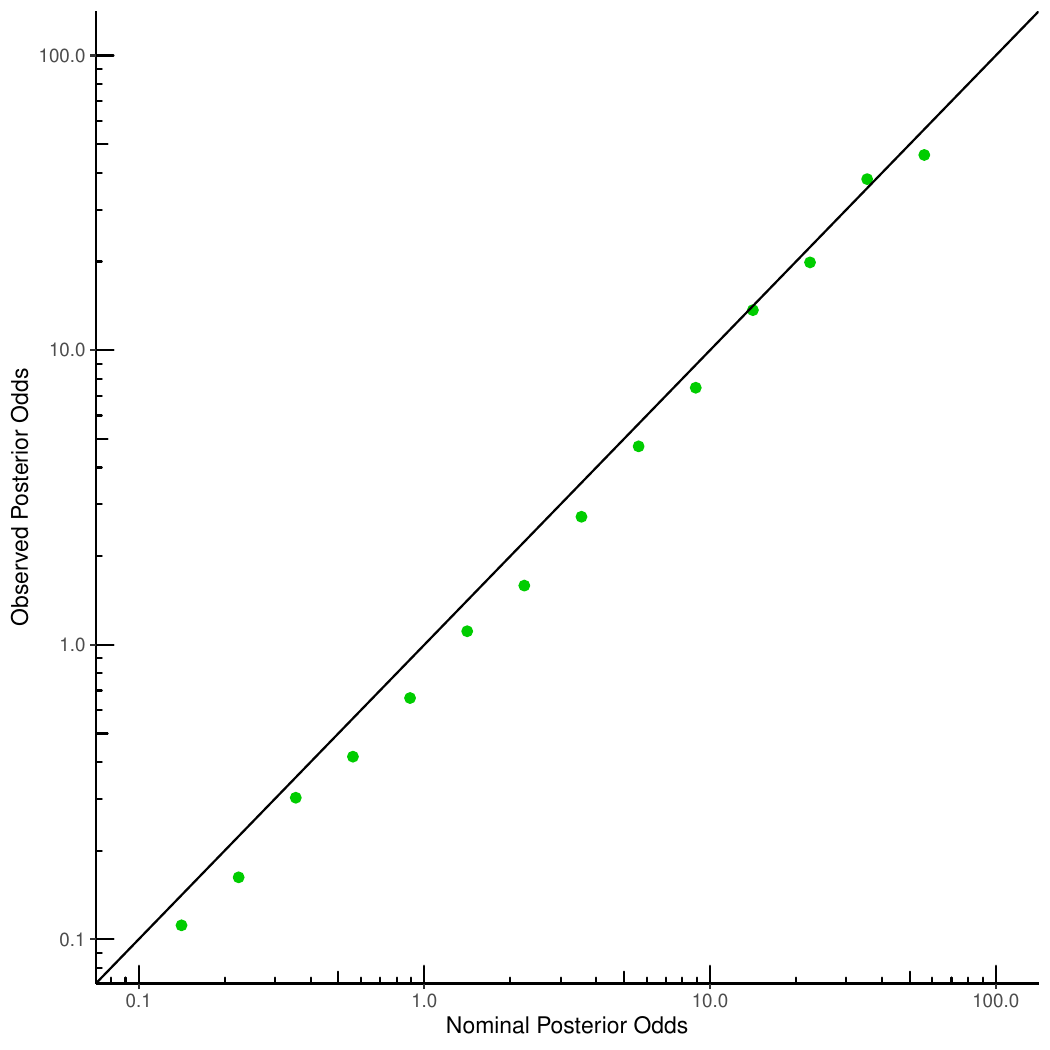}
		\caption{}
		\label{fig:Extra e}
	\end{subfigure}
	\hfill
	\begin{subfigure}[b]{0.4\textwidth}
		\includegraphics[width=\textwidth]{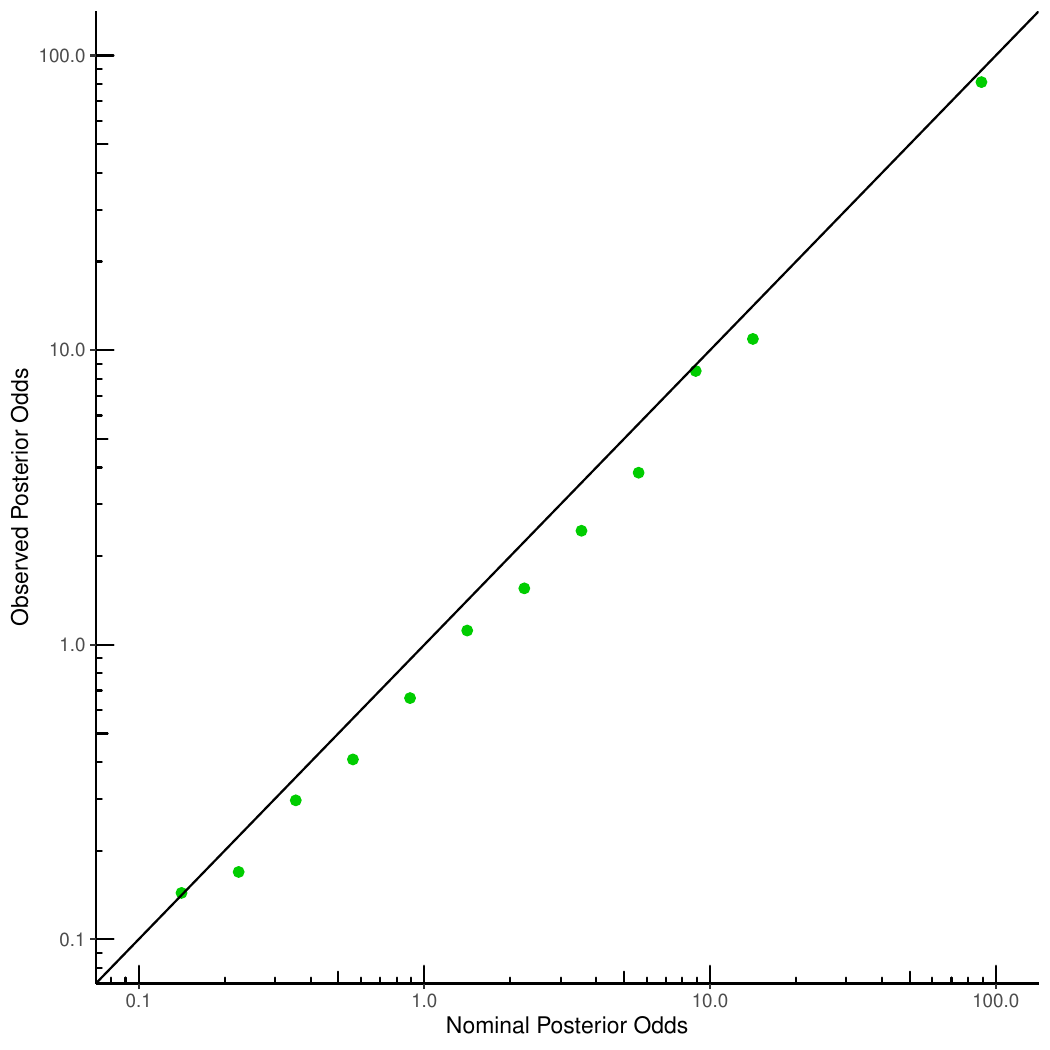}
		\caption{}
		\label{fig:Extra f}
	\end{subfigure}
	\caption{The observed posterior odds as a function of the nominal posterior odds, from $20.000$ replicate experiments. (a) Contingency table experiment, without optional stopping. (b) Contingency table experiment, with optional stopping.  (c) Subjective Bayesian version of the experiment in a. (d) Subjective Bayesian version of the experiment in b.}\label{fig:plaatjesExtra}
\end{figure}

We now revisit the example, but we change the proportions under both
hypotheses and survey only $50$ men and women, and we use a joint multinomial sampling scheme (the grand total, $n$, is fixed, but the number of men or women is not\footnote{Many thanks to Jorge Tendeiro for pointing out a mistake in in the sampling scheme in an earlier version of this technical report.}). Under $\H_0$,
$70\%$ of both men and women vote for the Democrats, and under $\H_1$,
$65\%$ of the men and $75\%$ of the women do. We repeat exactly the
same experiment (without optional stopping), and we see the resulting
plot in Figure~\ref{fig:Extra a}. We see that the relationship between
the observed and nominal posterior odds looks linear, but the slope is
off. If we repeat the same experiment with optional stopping, we see
in Figure~\ref{fig:Extra b} that additionally the linear association
is missing.\\

We do note that the objective priors used in the default Bayes Factor test for contingency tables are proper, so we are able to sample from them. In Figure~\ref{fig:Extra e} we see what happens if we do exactly the same experiment as in Figure~\ref{fig:Extra a}, but sampled from the prior: we see the observed posterior odds plotted against the nominal posterior odds, and the points lie approximately on the identity line, in contrast with Figure~\ref{fig:Extra a}. Furthermore, we performed the same experiment as in Figure~\ref{fig:Extra b} in this subjective Bayesian way, and we see that (in Rouder's terminology) the interpretation of the posterior odds holds with optional stopping in Figure~\ref{fig:Extra f}. As said, we do not think this kind of sampling is very meaningful in default prior context; we just add the experiment to show that invariably, {\em if\/} one can and wants to sample from priors, then Rouder's conclusions do hold. 

\paragraph{Subjective vs.\ Objective Interpretation}
In their original paper, \cite{Dickey1974} (GD) give a {\em
	subjective\/} interpretation to their priors. These priors depend on
the sampling scheme, i.e.\ on whether the grand total, and/or one or
both of the marginals are known or set by the experimenter in advance.
At first sight, this seems to be at odds with the fact that, with
subjective priors, Bayesian procedures do not depend on the stopping
rule used, as we pointed out in Section~\ref{sec:other}. However,
closer inspection reveals that if one follows the method under their
subjective interpretation, then the posterior indeed would not depend
on the sampling scheme. How is this possible? To see this, note that
GD do not model their data as coming in sequentially, but rather they
consider a fixed, single datum $D= (N_1, \ldots, N_4)$ consisting of
the four entries in the contingency table (see e.g.\ Table~\ref{tab:cont table general} below). The different versions of
their model and prior are then arrived at by calculating, for example,
${\mathbb P}(D \mid \H_0)$ for the case that no information about the
design is given, and ${\mathbb P}(D \mid \H_0, n)$ (where $n = N_1+
N_2 + N_3 + N_4)$ for the case that the grand total (sample size) $n$
is determined in the experiment design. In every case, the posterior
odds $\postodds{D}$ will remain the same; for they require the {\em
	prior\/} to be used when $n$ is given, ${\mathbb P}(\H_0 \mid n)$, to
be arrived at by conditioning the original prior ${\mathbb P}(\H_0)$ on
the grand total $n$. In particular, this means that a truly subjective
Bayesian who follows the GD model would have ${\mathbb P}(\H_0 \mid n)
\neq {\mathbb P}(\H_0)$, and could thus not use a $(1/2,1/2)$
`uninformative' prior on $(\H_0,\H_1)$ both when the grand total is
known in advance and when it is not. In other words, the posterior is
{\em not\/} affected by the sampling scheme, but the {\em prior\/} is.

\paragraph{Details of the experiments}\label{sec:details experiments cont tables}
For Example $4$ above, we used the function $\mathtt{contingencyTableBF}$. This function gives the user the option to choose between four different so called sampling schemes, implementing the \emph{Default Gunel and Dickey Bayes Factors for Contingency Tables} of \citet{Jamil2016}. Which of the four options to use, depends on which covariates in the contingency table are to be treated as fixed or as random, depending on the design of the experiment.

\begin{table}[h]
	\begin{center}
		\begin{tabular}{ c | c c | c }
			& $0$ & $1$ & sum\\
			\hline
			$0$ & $n_1-k_1$ & $n_2-k_2$ & $n-k$\\
			$1$ & $k_1$ & $k_2$ & $k$ \\
			\hline
			sum & $n_1$ & $n_2$ & $n$
		\end{tabular}
		\caption{$2$x$2$ contingency table; the four entries correspond to the numbers $N_1, N_2, \ldots, N_4$ above. \label{tab:cont table general}}
	\end{center}
\end{table}
In the first sampling scheme, none of the cell counts in the
contingency table are considered fixed, and the assumption is made
that each cell count is Poisson distributed. The default prior for
this scheme is a conjugate gamma prior on the Poisson rate parameter,
with hyperparameters suggested by Gunel and Dickey. We use this
sampling scheme for our first experiment in
Section~\ref{sec:contingency table exp.}, but as we noted in our
discussion in the same section, the question of `what is the actual
sampling scheme' and hence `what is the right default prior' for the
type of experiment we do --- the same experiment with and without
optional stopping --- is really impossible to answer. Thus, we
repeated the experiment with other (combinations of) sampling schemes,
in all cases obtaining similar results. Indeed, when we perform the
experiment without optional stopping, we sample a fixed number of men
and women, whereupon one margin $(n_1,n_2)$ and the grand total $(n)$ is fixed. For our
second example (Figure~\ref{fig:Extra a} and~\ref{fig:Extra b}) we used the prior advocated for the sampling scheme in which the grand total
($n$ in Table~\ref{tab:cont table general}) is fixed. Under this
sampling scheme, the cell counts are assumed to be jointly multinomial
distributed, and a Dirichlet conjugate distribution with the suggested
parameters \citep{Jamil2016} is used as prior, which in our case
amounts to a uniform prior on the Bernoulli parameter $\theta$;
see \citet{Jamil2016} for details. Again, using instead one of the priors advocated for one of the other sampling schemes leads to very similar results.

\end{document}